\documentclass[twocolumn,aps,pra,floatfix]{revtex4}

\usepackage{graphicx}
\usepackage[usenames,dvipsnames]{color}
\usepackage{dcolumn}
\usepackage{bm}
\usepackage{mathbbol}
\usepackage{amsmath,amssymb}  
\usepackage[latin9]{inputenc} 

\usepackage[sans]{dsfont}
\usepackage[mathscr]{euscript}

\usepackage{amssymb}          
\usepackage{amsmath}          

\usepackage{enumerate}
\usepackage[usenames,dvipsnames]{xcolor}
\usepackage[hidelinks]{hyperref}

\newtheorem{proposition}{Proposition}

\newtheorem{cor}{Corollary}
\newtheorem{lemma}{Lemma}

\newcommand{\proof}{\noindent {\em Proof. }}

\newcommand{\ket}[1]{|#1\rangle}
\newcommand{\bra}[1]{\langle #1|}
\newcommand{\braket}[2]{\langle #1|#2\rangle}

\newcommand{\Hi}{\mathcal{H}}

\newcommand{\Tr}{\mathrm{{Tr}}}
\newcommand{\supp}{\rm{supp}}

\newcommand{\beq}{\begin{equation}}
\newcommand{\eeq}{\end{equation}}
\newcommand{\beqa}{\begin{eqnarray}}
\newcommand{\eeqa}{\end{eqnarray}}
\newcommand{\beqan}{\begin{eqnarray*}}
\newcommand{\eeqan}{\end{eqnarray*}}

\newcommand{\C}{\mathbb{C}}

\newcommand{\qed}{\hfill $\Box$ \vskip 2ex}

\renewcommand{\ker}{{\rm ker}}
\newcommand{\spn}{{\rm span}}

\newcommand{\tr}{{\rm Tr}}

\newcommand{\HH}{{\mathfrak H}(\Hi)}

\newcommand{\BB}{{\mathfrak B}(\Hi)}

\newcommand{\HHs}[1]{{\mathfrak H}(\Hi_{#1})}
\newcommand{\Hf}{{\mathfrak H}}
\newcommand{\ft}{\color{black}}
\newcommand{\ftr}{\color{black}}
\newcommand{\ic}{\color{black}}
\begin{document}

\title{{Decompositions of Hilbert Spaces, Stability Analysis and Convergence Probabilities\\ for Discrete-Time Quantum Dynamical Semigroups}}

\author{Giuseppe Ilario Cirillo}
\email{giuseppeilario.cirillo@studenti.unipd.it}
\affiliation{Dipartimento di Ingegneria dell'Informazione,
Universit\`a di Padova, via Gradenigo 6/B, 35131 Padova, Italy}
\author{Francesco Ticozzi}
\email{ticozzi@dei.unipd.it}
\affiliation{Dipartimento di Ingegneria dell'Informazione,
Universit\`a di Padova, via Gradenigo 6/B, 35131 Padova, Italy, and Department of Physics and Astronomy, Dartmouth College, 6127 Wilder Laboratory, Hanover, NH 03755, USA}

\begin{abstract}            
We investigate convergence properties of discrete-time semigroup quantum dynamics, including asymptotic stability, probability and speed of convergence to 
pure states and subspaces. These properties are of interest in both the analysis of uncontrolled evolutions and the engineering of controlled dynamics for quantum information processing. Our results include two Hilbert space decompositions that allow for deciding the stability of the subspace of interest and for estimating of the speed of convergence,
{\ic as well as a formula to obtain the limit probability distribution for a set of orthogonal invariant subspaces.}
\end{abstract}


\maketitle

\section{Introduction}

Completely-Positive and Trace-Preserving (CPTP) maps on operator spaces have long been studied: they gained a central role in quantum open--system theory \cite{davies,kraus}, especially in the operator-algebra formualtion \cite{bratteli}, and more recently in quantum information theory and applications \cite{nielsen-chuang}. In essence, CPTP maps represent the most general evolution maps for the statistical description of quantum systems and thus are the main tool for describing the action of noise and the environment in quantum dynamical systems as well as quantum information processing protocols. In the control of quantum systems, open-system dynamics are needed whenever noise or measurements enter the model: CPTP evolutions are then associated to controlled open systems, in open or closed loop, and filtering equations \cite{altafini-tutorial}.

In terms of the qualitative analysis of the dynamics, the main difference between CPTP maps and unitary (or {\em coherent}) evolutions, is that the former allow for contractive dynamics. Two types of dynamical effects associated with CPTP dynamics are typically studied for uncontrolled dynamics: {\em decoherence} and {\em ergodicity}. {Decoherent} dynamics describe the decay of the quantum correlations between orthogonal states of a {\em preferred basis} and have been one of the main focuses of research concerned the action of environmental noise on quantum systems \cite{zurek-decoherence}. However, if we are interested in the engineering of quantum systems for control and information processing protocols, it is typically more interesting to know if we have convergence towards a target unique state, often pure and possibly entangled, or to a subspace \cite{burgarth-mixing, reeb-cutoff, lidar-DFS, ticozzi-QDS, ticozzi-markovian, bolognani-arxiv, ticozzi-stochastic,mirrahimi-stabilization, albertini-feedback, yamamoto-twospin, ticozzi-QL, ticozzi-QIC, barreiro, barreiro2}.

In this work, we develop analysis tools that address the issues emerging in problems of dynamical engineering. We consider {\ic an {\em invariant subspace} }
of the Hilbert space associated to the quantum system of interest, which we assume throughout the paper {\em to be finite-dimensional.} More precisely, we assume that the set of operators with support limited with that subspace is invariant. The class of dynamical models we focus on is associated to iteration of a given CPTP map, i.e. a discrete-time quantum dynamical semigroup in the language of \cite{alicki-lendi,bolognani-arxiv}. We thus investigate {\ic the asymptotic behavior with respect to this subspace, i.e. {\em whether the evolution converges towards it}}, 
 {\em how fast the convergence is} and, 
 {\ft {\em the asymptotic probability distribution} with respect to smaller invariant subspaces that are comprised in it.} Clearly, convergence to a pure state, be it entangled or not, can be seen as a particular case of convergence to a (one-dimensional) subspace.  

Some of these problems have been partially addressed for continuous-time CPTP semigroups, notably in the work of Baumgartner {et al.} \cite{baumgartner-1,baumgartner-2} and Ticozzi {\em et al.} \cite{ticozzi-NV}, using linear-algebraic approaches. 
We here build on these ideas and develop the analysis directly in the discrete-time case. This is more general, as not all discrete-time semigroup can be seen as ``sampled'' continuous-time semigroups \cite{wolf-dividing}, and has its own peculiarities. 

We employ an approach based on Perron-Frobenious theory that allows to follow the same path taken in the study of the analogous classical Markov chain properties. We first propose two alternative methods, both based on suitable Hilbert space decompositions in ``transient subspaces'', to analyse convergence. The first is based on the spectral structure of the linear CPTP map, while the second on the ``dissipative links'' between the subspaces. They both provide constructive tests for the attractivity of {\ic an invariant subspace,} 
and ways to estimate the asymptotic speed of convergence to it.  Both the approaches build on existing ergodicity results and representations of CPTP dynamics \cite{wolf-notes, bolognani-arxiv}. In the spirit of \cite{ticozzi-NV}, we do not assume that the target subspace collects all the invariant sets of states, so the algorithms can be used as tests for the asymptotic stability of {\ic any sets}.

{\ic In the case that the considered subspace} is not asymptotically stable, they both allow to extend {\ic it to an attractive one.} 
The dissipation-induced decomposition, which is initially presented in Schr\"odinger's picture, is further developed and clarified in the dual picture.  Finally, we propose some explicit formulas to compute convergence to invariant, yet non attractive subspaces. These build on similar results for Markov chains, and suitable map decompositions.

Sections \ref{sec:preliminaries} and \ref{sec:invariant} presents the key definitions, notations and some basic results on CPTP maps and invariant structures. The two main decompositions and their properties are described in Sections \ref{sec:nested} and \ref{sec:DID}, respectively. Section \ref{sec:asProb} contains the derivations of the formulas for the asymptotic convergence probabilities, while an illustrative example is presented in Section \ref{sec:example}.

\section{Preliminaries}\label{sec:preliminaries}
\subsection{Definitions and Notations}
Consider a finite-dimensional Hilbert space $\Hi\sim\C^d.$ Let $\mathfrak{B}(\mathcal{H})$ be the (complex) space of linear operators on  $\Hi$, $\mathfrak{H}(\mathcal{H})$ the (real) space of hermitian operators{\ic, and  $\mathfrak{H}^+(\mathcal{H})$ the cone of Positive Semi-Definite (PSD) operators}.
A linear map $T:\mathfrak{H}(\mathcal{H}) \to \mathfrak{H}(\mathcal{H})$ is said to be CPTP if it is:
\begin{enumerate}
\item \emph{Trace preserving} {\color{black} (TP)}:
\begin{equation}
\forall \,\, A \in \mathfrak{H}(\mathcal{H}): \,\,\,\,\, \tr(T(A))=\tr(A).
\end{equation}
\item \emph{Completely positive} (CP):
$\forall n\in \mathbb{N}$, $T\otimes id_n$ is positive, where $id_n$ is the identity map on the operators of an Hilbert space of dimension $n$.
\end{enumerate}

All physically-admissible evolutions are CPTP \cite{nielsen-chuang}. Consider the set of density operators on $\Hi$, or quantum states, $\mathfrak{D}(\mathcal{H})$, namely the set of self-adjoint, positive semi-definite operators with trace one. Clearly, if $T$ is a CPTP map, it leaves $\mathfrak{D}(\mathcal{H})$ invariant. Hence, $T$ can be seen as the generator of a (discrete) \emph{Quantum Dynamical Semigroup} (QDS) in Schr\"odinger picture by considering iterated applications of the map:
\begin{equation}
\rho(n+1)=T(\rho(n)),
\end{equation}
for any $\rho(0)\in \mathfrak{D}(\mathcal{H}).$

The Stinespring-Kraus representation theorem \cite{kraus} ensures that a linear map is CP (not necessarily CPTP) if and only if it admits an Operator-Sum Representation (OSR):
\begin{equation}\label{eq:Kraus}
T(\rho)=\sum\limits_{k=1}^{K}M_k\rho M_{k}^{\dagger},
\end{equation}
for some $M_k \in \mathfrak{B}(\mathcal{H}),$ where $M^\dagger$ is the adjoint of $M$. For $T$ to be also CPTP it is easy to see, using the cyclic property of the trace, that it is necessary and sufficient to require that:
\begin{equation}\label{eq:unital}
\sum\limits_{k=1}^{K} M_{k}^{\dagger}M_k=I,
\end{equation}
where $I$ is the identity operator on $\Hi.$

{\ft
As for any linear map $T$, it is possible to consider its Jordan decomposition:
\begin{equation}\label{eq:jordDec}
T=\sum\limits_{\lambda_k \in \lambda(T)} \lambda_k P_k + N_k,
\end{equation}
where $P_k$ is the projection onto the generalized eigenspace in $\mathfrak{H}(\mathcal{H})$ relative to the eigenvalue $\lambda_k$, and $N_k$ is nilpotent with index $d_k$ ($N_k^{d_k}=0$). The eigenvalues coincides with the roots of the {\em characteristic polynomial} of $T$: 
\begin{equation}
\Delta_T(\lambda)=\det(\lambda I -T).
\end{equation}
The nilpotent index $d_k$ is determined by the multiplicities of $\lambda_k$ as a root of the {\em minimal polynomial} of $T$, $\psi_T(\lambda)$, i.e. the one with minimal degree among the polynomials $p(\lambda)$ such that $p(T)=0$.
Moreover, the positivity of the map allows for a generalization of Perron-Frobenius theory \cite{wolf-notes}, which guarantees that the set of eigenvalue contains its spectral radius.
 

The space $\HH$ is a Hilbert-Space if endowed with the Hilbert-Schmidt inner product:
\begin{equation}\label{eq:HSpro}
\braket{A}{B}=\tr(AB), \,\,\, A,B \in \HH.
\end{equation} 
Let $T^*$ be the dual of $T$ with respect to the above inner product. Given an OSR for $T$, it follows from~\eqref{eq:HSpro} that $T^*$ acts on an element $A\in\HH$ by:
\begin{equation}
T^*(A)=\sum_k M_k^\dagger A M_k.\end{equation}
Thus, the dual of a CP map is still CP. However, it is immediate to show that the dual of a CPTP map does not need to be TP, but it must be {\em unital}, since \eqref{eq:unital} holds.

\subsection{Block representations}
To our aim, it is useful to introduce appropriate ``block'' representations of maps and operators with respect to a decomposition of the underlying Hilbert space. Consider a decomposition of $\Hi$ into two orthogonal subspaces:
\begin{equation}\label{eq:spaceDec}
\mathcal{H}=\mathcal{H}_S\oplus\mathcal{H}_R.
\end{equation}
If we choose a basis:
\begin{equation}
 \{|\varphi_l\rangle\}=\{|\phi_l^S\rangle\} \cup \{|\phi_l^R\rangle\},
\end{equation}
where $\{|\phi_l^S\rangle\}$ is a basis for $\mathcal{H}_S$ and $\{|\phi_l^R\rangle\}$ for $\mathcal{H}_R$, a block structure is induced on any matrix representing an element $X\in\BB$:
\begin{equation}\label{eq:indBlockForm}
 X =\left[\begin{array}{cc}X_S & X_P \\ X_Q & X_R \end{array}\right],
\end{equation}
and in particular on the matrices $M_k$ appearing in an OSR of $T$.

{\ftr Building on this decomposition, we can see the space of hermitian operators $\mathfrak{H}(\mathcal{H})$  as a sum of three orthogonal operator subspaces:}
\begin{equation}\label{eq:hermDec}
\begin{aligned}
\mathfrak{H}(\mathcal{H})& = \mathfrak{H}_S \oplus \mathfrak{H}_{SR} \oplus \mathfrak{H}_R,\\
\mathfrak{H}_S& = \left\{ \rho \in \mathfrak{H}(\mathcal{H}) : \rho=\left[\begin{array}{cc}\rho_S & 0 \\ 0 & 0 \end{array}\right] \right\},\\
\mathfrak{H}_{SR}& = \left\{ \rho \in \mathfrak{H}(\mathcal{H}) : \rho=\left[\begin{array}{cc} 0 & \rho_P \\ \rho_P^\dagger & 0 \end{array}\right] \right\},\\
\mathfrak{H}_R& = \left\{ \rho \in \mathfrak{H}(\mathcal{H}) : \rho=\left[\begin{array}{cc}\ 0 & 0 \\ 0 & \rho_R \end{array}\right] \right\}.
\end{aligned}
\end{equation}
We shall use the notation $\mathfrak{H}_S^+$ for the PSD elements in $\mathfrak{H}_S$, and similarly for $\mathfrak{H}_R^+$.
From~\eqref{eq:spaceDec} and \eqref{eq:hermDec} it is possible to construct three ``reduced'' linear maps:
\beqa
T_S&:&\Hf_S \rightarrow \Hf_S \nonumber \\ 
T_S(A_S)&=&\sum \limits_k M_{k,S} A_S M_{k,S}^\dagger,\label{eq:TsDef} \\
T_R&:&\Hf_R \rightarrow \Hf_R \nonumber \\ 
T_R(A_R)&=&\sum \limits_k M_{k,R} A_R M_{k,R}^\dagger,\label{eq:TrDef}\\
T_{SR}&:&\Hf_R \rightarrow \Hf_S \nonumber \\ 
T_{SR}(A_R)&=&\sum \limits_k M_{k,P} A_R M_{k,P}^\dagger. \label{eq:TsrDef}
\eeqa

\noindent Notice that these maps can be defined without explicitly referring to an OSR of $T$, by considering e.g. 
\beq
T_{SR}(A_R) = \Pi_S T(A_R) \Pi_S, \,\,\,\,\,\,\, A_R \in \Hf_R,
\eeq
where $\Pi_S$ is the orthogonal projection on $\Hi_S$. 

\section{Invariant and Asymptotically Stable Subspaces}\label{sec:invariant}
\subsection{Definitions}

Given an $X\in\BB$, its support is defined as:
\begin{equation}
\supp(X)=(\ker(X))^\perp,
\end{equation}
and coincides with its range if $X\in\HH$. Moreover, given {\em a set} of operators $W\subset\HH$, its support can be defined as the minimal subspace of $\Hi$ that contains the supports of all the elements in $W$:
\begin{equation}
\supp(W)=\bigvee\limits_{\eta\in W}\supp(\eta),
\end{equation}
where $\vee$ stands for the sum between subspaces.

Studying how the support of an operator evolves, along with a decomposition like~\eqref{eq:spaceDec}, is instrumental to study the properties of {\em invariant subspaces}: $\Hi_S$ is said invariant if any trajectory starting from an operator with support in it, has its support in $\Hi_S$ for all times, or equivalently:
\begin{equation}
\supp(\rho)\subset\Hi_S \Rightarrow \supp(T(\rho))\subset\Hi_S. 
\end{equation}

A particular class of invariant subspaces, is that one of {\em Globally Asymptotically Stable} (GAS). {\ftr These subspaces are also globally attractive:  an invariant $\Hi_S$ is said to be GAS if:
\begin{equation}
\lim\limits_{n \to \infty} \|T^n(\rho)-\Pi_S T^n(\rho)\Pi_S\| = 0,
\end{equation}
\[\forall \rho \in \mathfrak{D}(\mathcal{H}),\]
{\ic where $\Pi_S$ is the projection onto the subspace.} Namely, they are subspaces to which the evolution converges.}

\subsection{Invariant Faces for CPTP maps}

We begin by recalling some basic properties of invariant sets for CPTP maps acting on the cone {\ic $\HH^+$}. 
In particular we shall focus on {\em faces} of the cone.

Given a CP map $T$, a face $F$ is invariant if
\beq
\forall M \in F, \,\,\,\, T(M) \in F. \nonumber
\eeq
It is well-known that any face corresponds to the set of PSD operators with support on a subspace. Moreover, the correspondence between faces and subspaces is a lattice isomorphism \cite{hill1987cone}. Using this correspondence, {\color{black} it is easy to verify that a face is invariant if and only if the corresponding subspace is so}. The next Lemma shows that in fact to any invariant set corresponds an invariant face.
\begin{lemma}
\label{lem:invSet}
Let $T$ be a positive map with $W\subset \mathfrak{H}^+(\mathcal{H})$ an {\em invariant} set. If $X \in \mathfrak{H}^+(H)$ is such that $\supp(X) \subset \supp(W)$ then $\supp(T(X))\subset \supp(W)$.
\end{lemma}

$\\ $
\proof
Let $\widetilde{W}$ be the convex {\color{black} hull} of $W$, so $\supp(\widetilde{W})=\supp(W)$, $\widetilde{W}$ is invariant and contains an element $A$ such that $\supp(\widetilde{W})=\supp(A)$. Thus $\supp(T(A))\subset\supp(A)$ since $T(A) \in \widetilde{W}$. By $\supp(X) \subset \supp(W)$ there exists a constant $c>0$ such that $cA \geq X$ so $T(cA) \geq T(X)$ which implies $\supp(T(X)) \subset \supp(T(cA))$.
\qed

The next property characterizes the invariance in terms of the OSR of a CP map $T$, and slightly extends some results of \cite{bolognani-arxiv}.
\begin{proposition}
\label{prop:genInva}
Suppose $T$ is a CP map on $\HH$ and $\Hi=\Hi_S\oplus\Hi_R$ then
\begin{enumerate}[i)]
\item $\Hi_S$ is invariant if and only if in any OSR for $T$ the matrices $M_k$ have the block structure:
\beq
\label{eq:invGenKraus}
M_k=\left[\begin{array}{cc}
	M_{k,S} & M_{k,P} \\
	 0      & M_{k,R}
\end{array}
\right].
\eeq
\item $\Hi_S$ is invariant if and only if $\Hf_{S} \oplus \Hf_{SR}$ is invariant under the action of $T$.
\end{enumerate}
\end{proposition}
\proof 
\begin{enumerate}[i)]
\item Fix an OSR for $T$ and let $A \in \Hf^+_S$
\beq
A=\left[\begin{array}{cc}
A_S & 0 \\
0 & 0
\end{array}\right].
\eeq
By applying $T$ we obtain:
\beq
T(A)=\left[\begin{array}{cc}
\sum\limits_k M_{k,S}A_S M_{k,S}^\dagger & \sum\limits_k M_{k,S}A_S M_{k,Q}^\dagger \\
\sum\limits_k M_{k,Q}A_S M_{k,S}^\dagger & \sum\limits_k M_{k,Q}A_S M_{k,Q}^\dagger 
\end{array}\right].
\eeq
If $\Hi_S$ is invariant, we must have ${\sum\limits_k M_{k,Q}A_S M_{k,Q}^\dagger=0}$ for any choice of $A_S.$ Since it is a sum of positive elements, all summands must be zero, irrespective of $A_S.$ This implies that $M_{k,Q}=0$ for any $k$.
Conversely, if $M_{k,Q}=0$ for any $k$ then for any $A \in \Hf^+_S$, $T(A)\in \Hf^+_S$ {\color{black} since all the others blocks} are zero as well, and $\Hi_S$ is invariant.

\item If $\Hf_{S} \oplus \Hf_{SR}$ is invariant and $T$ is positive, then the intersection $\Hf^+_S = \Hf_{S} \oplus \Hf_{SR} \cap \Hf^+$ is invariant as well.
On the other hand, if $\Hi_S$ is invariant consider a generic $A \in \Hf_{S} \oplus \Hf_{SR}$, $A=\left[\begin{array}{cc}
A_S & A_P \\
A_Q & 0
\end{array}\right]$. If we now apply $T$ and use the OSR structure we derived in \eqref{eq:invGenKraus} we obtain {\color{black} the following structure:
\beq
T(A)=\left[\begin{array}{cc}
* & * \\
* & 0
\end{array}\right],
\eeq}
and so $\Hf_{S} \oplus \Hf_{SR}$ is invariant.
\end{enumerate}
\qed

{\color{black}
Using this result we can infer some properties for the reduced maps~\eqref{eq:TsDef} and~\eqref{eq:TrDef}. Assuming that $\Hi_S$ is invariant, $T_S$ is just the restriction of $T$ to $\Hf_S$, while $T_R$ is its effect (pinching) on $\Hf_R$. Then, from the proposition it follows that that the product of the characteristic polynomials of $T_S$ and $T_R$ divide the characteristic polynomial of $T$, while their minimal polynomials divide that one of $T$. In particular the eigenvalue of $T_S$ and $T_R$ are eigenvalue of $T$ too and, moreover, the index of an eigenvalue of $T_S$ or $T_R$ is lesser or equal than the index of the same eigenvalue with respect to $T$.}

\section{Nested-Face Decomposition}\label{sec:nested}
\subsection{Preview of the key results}
We are now ready to focus on the main objective of this paper: studying the convergence features of an iterated CPTP map $T$ by identifying ``transient'' faces of {\ic the cone of PSD operators. }
{\ftr The first approach proposed construct an increasing sequence of {\em nested} faces of PSD operators, each included in all the following ones. For this reason, we will name it Nested Face Decomposition (NFD). Starting from an invariant subspace, such sequence is associated to a sequence of subspaces $\{\Hi_{S_i}\}$, and it is iteratively built relying on the positivity and spectral properties of the map. 

The idea is the following: at each step, the part of the sequence already constructed determines a subspace $\Hi_{S_i}$, and the total space of the system is decomposed in the latter and its orthogonal complement $\Hi_{R_{i}}=\Hi_{S_i}^\perp$. 
The next face/subspace $\Hi_{S_{i+1}}$ is obtained as follows: we start by reducing the map $T$ to the orthogonal complement $\Hi_{S_i}^\perp$, and finding the operator eigenspace $D_i$ associate to its slowest eigenvalue (its spectral radius). By adding the support of $D_i,$ call it $\Hi_{T_i},$ to $\Hi_{S_i}$, a new element of the nested sequence is obtained:
\[\Hi_{S_{i+1}}=\Hi_{S_{i}}+\Hi_{T_{i}}, \quad \Hi_{S_{i}}\subseteq\Hi_{S_{i+1}}.\]
At the end, the last face corresponds to the whole Hilbert space, which now can be decomposed as:
\[\Hi=\Hi_S\oplus\left(\bigoplus_i\Hi_{T_i}\right).\]
We next describe an extremely simple example that can be studied avoiding mathematical complications, and illustrates the basic principles of the decomposition. A more complete application of our results will be described in Section \ref{sec:example}.

{\em Toy model: 3-level system for NFD.} Consider a three level system on a Hilbert space spanned by orthonormal vectors $\{\ket{1},\ket{2},\ket{3}\}$, and assume on each time interval of length $T$ it can undergo the following processes: populations on levels  $\ket{1},\ket{2}$ are swapped with probability $\gamma_1$, and left as they are with probability $\gamma_0$; 
level $\ket{3}$ decays to level $\ket{1}$ with probability $\gamma_2$, while leaving the other two levels untouched.
Assume these processes can be represented, with respect to the reference basis, by the following OSR:
\[M_0=\sqrt{\gamma_0}\left[ \begin{array}{ccc}
1  & 0 & 0 \\
0 & 1 & 0 \\
0 & 0 & 1
\end{array}
\right];
M_1=\sqrt{\gamma_1}\left[ \begin{array}{ccc}
0  & 1 & 0 \\
1 & 0 & 0 \\
0 & 0 & 1
\end{array}
\right];\]
\beq
M_2=\sqrt{\gamma_2}\left[ \begin{array}{ccc}
0 & 0 & 1 \\
0 & 0 & 0 \\
0 & 0 & 0
\end{array}
\right].\label{OSR-ex}\eeq
Assuming $\sum_i\gamma_i=1$ the map is TP. It is easy to see that the pure state $\rho_S=1/2(\ket{1}+\ket{2})(\bra{1}+\bra{2})$ is invariant for the dynamics. For the purpose of illustration, we can thus try to use the ideas sketched above to analyze if it is GAS, and the convergence properties of the map. Let $\Hi_{S_1}=\spn\{\frac{1}{2}(\ket{1}+\ket{2})\}.$ The positive map reduced to its complement, $\Hi_{R_1}= \spn\{\frac{1}{2}(\ket{1}-\ket{2}),\ket{3}\}$, has spectral radius $1,$ since $\rho_1=\frac{1}{2}(\ket{1}-\ket{2})(\bra{1}-\bra{2})$ is also fixed. Then we can add 
$\Hi_{T_1}=\spn\{\frac{1}{2}(\ket{1}-\ket{2})\}$ to $\Hi_{S_1}$ and obtain
$\Hi_{S_2}=\Hi_{S_1}\oplus\Hi_{S_2}.$ Now its complement $\Hi_{R_2}$ has spectral radius $1-\gamma_2<1,$ and the reduced map is just the multiplication by this scalar. So we have $\Hi_{S_2}$ is an enlargement of $\Hi_{S_1}$ that makes it GAS. \qed

In general, we shall prove that this construction assures that the reduced maps are characterized by {\em strictly decreasing} spectral radii, with corresponding eigenoperators having nontrivial support on the $\Hi_{T_{i}}$. This allows to estimate the asymptotic speed with which the probability decays on the subspaces of the sequence, and thus the speed of convergence to the smaller faces. Finally, this allows us to decide if the starting subspace is GAS or not: it is necessary and sufficient that the spectral radius of the first reduced map is strictly lesser than one.}

\subsection{Construction}
To start assume $\Hi_{S_1}=\Hi_S,$ {\ic an} initial invariant subspace, and $\Hi_{R_1}=\Hi_R=\Hi\ominus\Hi_S$.
The general construction step of $\Hi_{R_{i+1}}$ and $\Hi_{S_{i+1}}$ from $\mathcal{H}_{R_{i}}$ and $\mathcal{H}_{S_{i}}$ can be carried out as follows:
\begin{enumerate}
\item Define

\beqa
\label{eq:Di}
D_i&:=&\ker((T_{R_i}-\sigma(T_{R_i})I)^{d_i^2}) \cap \Hf^+_{R_i}\\
\Hi_{T_i}&:=&\supp(D_i)\label{eq:Ht}
\eeqa
where $d_i=\dim(\Hi_{R_i})$, and $\sigma(T_{R_i})$ is the spectral radius of $T_{R_i}$. Since $T_{R_i}$ is CP, $D_i$ contains elements different from zero \cite{wolf-notes}. Furhtermore $D_i$ is the intersection of two invariant sets for $T_{R_i}$, so it is invariant. By Lemma \ref{lem:invSet} $\Hi_{T_i}$ is then invariant as well;
\item Call $\Hi_{S_{i+1}}=\Hi_{S_{i}}\oplus\Hi_{T_i}$. Notice that by construction this is an invariant subspace for $T$;
\item If $\Hi_{T_i}=\Hi_{R_i}$ the construction is ended, otherwise define $\Hi_{R_{i+1}}=\Hi_{S_{i+1}}^\perp$, and iterate.
\end{enumerate}
Since at any step $\Hi_{T_i}$ is not the zero space, $\Hi_{S_{i+1}}$ has dimension strictly larger than the dimension of $\Hi_{S_{i}}$, so in a finite number of steps its dimension must reach the dimension of $\Hi$ and the construction stops. When this is the case, we have obtained a chain of subspaces with the following properties:
\beqa
\Hi_{S_1}&=&\Hi_S,\label{eq:s1eqs}\\
\Hi_{S_i}&\subset & \Hi_{S_{i+1}}, \label{eq:inclusion}\\
\Hi_{S_N}&=&\Hi.
\eeqa
Alternatively, by using the $\Hi_{T_i}$ we introduced in Equation \eqref{eq:Ht} of the iterative construction, this decomposition can be rewritten as
\beq \label{eq:NFdec}
\Hi=\Hi_S\oplus\left(\bigoplus_i\Hi_{T_i}\right).
\eeq
It is then easy to see that, in a basis that reflect this structure, the matrices $M_k$ are in a block upper-triangular form, with the first diagonal block corresponding to $\Hi_S$ and each other diagonal block corresponding to one of the $\Hi_{T_i}$.

\subsection{NFD Properties}\label{sec:NFprop}
\begin{proposition}\label{prop:decRad}
If the chain \eqref{eq:inclusion} is constructed as above, then the spectral radii of the reduced $T_{R_i}$ on $\Hi_{R_i}=\Hi_{S_i}^\perp$ satisfy:
\beq
\label{eq:propDecRad}
\sigma(T_{R_i})>\sigma(T_{R_{i+1}}).
\eeq
\end{proposition}
{\ic \proof See appendix~\ref{ap:decRadProof}.}

This decomposition thus provides us with a nested sequence of faces to which the cone of PSD matrices asymptotically converge. In fact, if each of the $\Hi_{T_i}$ is characterized by a spectral radius strictly lesser than one, these subspaces tend to correspond to zero probability asymptotically.  The next proposition shows how this is naturally related to attractivity.
\begin{proposition}\label{prop:leastInvNF}
 The invariant subspace $\Hi_S$ is GAS if and only if $\sigma(T_{R_1})<1$. If this is not the case, $\Hi_{S_2}=\Hi_{S}\oplus\Hi_{T_1}$ is then the minimal GAS subspace containing $\Hi_S$.
\end{proposition}
\proof
If $\sigma(T_R)<1$ the support of any fixed points for $T$ is contained in $\Hi_S$ so by the results in \cite{bolognani-arxiv} $\Hi_S$ is GAS. If $\sigma(T_R)=1$ then $\sigma(T_{R_2})<1$ and $\Hi_{S_2}$ is GAS. Moreover there exist a density operator $\rho$ with $\supp(\rho)=\Hi_{T_i}$ and $T_R(\rho)=\rho$, where the last is a consequence that for $T$ 1 has only simple eigenvectors \citep{wolf-notes} and so {\color{black} the same holds for} $T_R$, by this any GAS subspace must contain $\Hi_{T_1}$.
\qed
 We conclude the section with two remarks. First, all the construction is based on the maps $T_{R_i},$ which do not depend on a particular OSR, but only on the (whole) map $T$. Secondarily, if $\mathcal{H}_S$ is taken to be the support of the {\em whole} fixed points subspace, the above construction is similar in principle to the decomposition in ``transient subspaces'' that has been proposed in \cite{baumgartner-2} for continuous time evolutions. However, beside being developed for discrete-time semigroups, our results differs from the above in the following aspects: (i) we allow for the initial subspaces to be {\em any} invariant subspace, a generalization that is of interest in many control and quantum information protection tasks \cite{ticozzi-QDS,ticozzi-markovian,ticozzi-NV,ticozzi-isometries,viola-generalnoise,lidar-DFS}; (ii) we investigate the structure and the spectral properties of the emerging $\mathcal{H}_{T_i}$; (iii) we use the latter as tool for {\em deciding} asymptotic stability of the subspace.

\section{DID Decomposition}\label{sec:DID}
\subsection{Preview of the key results}
{\ftr
The second decomposition we study is the discrete-time version of a decomposition first proposed in \cite{ticozzi-NV} for continuous quantum dynamical semigroups. The starting point is again {\ic an} invariant subspace $\Hi_S$, but in this case we construct a chain of subspaces that have to be ``crossed'' by the state trajectory in order to reach the invariant subspace $\Hi_S$, also in the form:
\[\Hi=\Hi_S\oplus\left(\bigoplus_i\Hi_{T_i}\right).\]
The idea underlying the construction is the following: as in the case of the nested-face decomposition, at each step of the construction we start with the subspaces of the chain have already been obtained, namely $\Hi_{S_i}=\Hi_S\oplus\Hi_{T_1}\oplus\ldots\Hi_{T_{i-1}}.$ The full space is then decomposed into $\Hi=\Hi_{S_i}\oplus \Hi_{R_i}$. The new element $\Hi_{T_i}$ is constructed as the part of $\Hi_{R_i}$ that is {\em dynamically connected} to the previous transient subspace $\Hi_{T_{i-1}},$ that is, the states with support on $\Hi_{T_i}$ will be mapped by $T$ into states that have support {\em also} on $\Hi_{T_{i-1}}$ 

The iterative construction ensures that $\Hi_{T_i}$ can be connected only to  the last $\Hi_{T_{i-1}},$ and allows for an approximate estimation of the rates at which the density operators with support on one of the $\Hi_{T_i}$ tend to flow towards $\Hi_S,$ following the chain of subspaces. Finally, it is easy to use the construction to determine stability of the initial $\Hi_{S}$ : if the last subspace is linked to the previous one, the whole chain is connected and we shall prove that this makes $\Hi_{S}$ GAS.

Let us briefly revisit the toy model we introduced in order to illustrate the NFD, and intuitevly show how the DID procedure works and differs from the latter.

{\em Toy model: 3-level system for DID.} Consider again the three level system on a Hilbert space spanned by orthonormal vectors $\{\ket{1},\ket{2},\ket{3}\}$, and assume on each time interval of length $T$ undergoes a CPTP evolution associated to OSR $\{M_0,M_1,M_2\}$ as in \eqref{OSR-ex}. Being the state $\rho_S=1/2(\ket{1}+\ket{2})(\bra{1}+\bra{2})$ invariant, we can start by the invariant subspace $\Hi_{S_1}=\spn\{\frac{1}{2}(\ket{1}+\ket{2})\}.$ The only part of the Hilbert space dynamically connected to $\Hi_{S_1}$ is $\Hi_{T_1}=\spn\{\ket{3}\},$ via the decay process associated to $M_2.$ So we can define $\Hi_{S_2}=\Hi_{S_1}\oplus\Hi_{T_1}$ and see how the remaining $\Hi_{R_2}=\spn\{\frac{1}{2}(\ket{1}-\ket{2})\}$ is conencted to the preceding. It is clear that while there is decay from $\Hi_{T_1}$ to $\Hi_{R_2},$ the viceversa is not true and $\Hi_{R_2}$ remains invariant. Thus, $\Hi_{S_1}$ cannot be GAS.  

So while $\Hi_{T_1}$ for the NFD was $\spn\{\frac{1}{2}(\ket{1}-\ket{2})\}$, being the support of the spectral radius eigenoperator of the reduced map, in this case it is $\spn\{\ket{2}\},$ collecting the part of the Hilbert space that is dynamically connected to the initial invariant subspace. \qed}

\subsection{Construction}

Fix an OSR  $\{M_k\}$ for $T$. To start, let $\Hi_{S_1}=\Hi_S,$ {\ic an} initial invariant subspace, and $\Hi_{R_1}=\Hi_{R}=\Hi_S^\perp.$
 We proceed iteratively: at each step we start from a decomposition of the form
\beq \label{eq:didBigDec}
\Hi=\Hi_{S_i}\oplus\Hi_{R_i},
\eeq
where 
\beq \label{eq:didHSiDec}
\Hi_{S_i}=\Hi_S\oplus\bigoplus\limits_{j=1}^{i-1}\Hi_{T_j}.
\eeq
Let $M_{k,P'_i}$ and $M_{k,R_i}$ be the $P$- and $R$-blocks in \eqref{eq:invGenKraus}, with respect to the decomposition \eqref{eq:didBigDec}. First $\Hi_{R_{i+1}}$ is defined as follows
\beq
\Hi_{R_{i+1}}=\bigcap _k \ker(M_{k,P'_i}).
\eeq
Three cases are possible:
\begin{enumerate}
\item for all $k$ $M_{k,P'_i}=0$ \emph{i.e.} $\Hi_{R_{i+1}}=\Hi_{R_{i}}$. In this case $\Hi_{T_{i}}=\Hi_{R_{i}}$, so that $\Hi_{S_{i+1}}=\Hi$, and the construction is terminated.
\item $\Hi_{R_{i+1}}=\{0\}$: in this case the construction is concluded. Again $\Hi_{T_{i}}=\Hi_{R_{i}}$.
\item If none of the above cases applies, we choose as $\Hi_{T_{i}}$ the orthogonal complement of $\Hi_{R_{i+1}}$ in $\Hi_{R_{i}}$
\beq
\Hi_{R_{i}}=\Hi_{T_{i}}\oplus\Hi_{R_{i+1}}.
\eeq
Since degenerate cases have been dealt separately, both $\Hi_{T_{i}}$ and $\Hi_{R_{i+1}}$ have positive dimension and then $\Hi_{R_{i+1}}$ has dimension strictly less than $\Hi_{R_{i}}$. Define $\Hi_{S_{i+1}}=\Hi_{S_i} \oplus \Hi_{T_i}$ and iterate.
\end{enumerate}

Notice that at any step the dimension of $\Hi_{R_{i}}$ strictly decreases, or the procedure is halted. Thus, in a finite number of steps the procedure terminates.
When the algorithm stops the subspace is decomposed in a direct sum of the form 
\beq
\label{eq:DIDdec}
\Hi=\Hi_S\oplus\bigoplus\limits_{j=1}^{N}\Hi_{T_j}.
\eeq
We shall call this the discrete-time {\em Dissipation-Induced Decomposition} (DID). {\ft In addition, we will call DID constructions concluded by case 2 above {\em successful}, while those terminated by case 1 unsuccessful. The reason will be clarified in Proposition \ref{prop:didgas}, where we show that case 2 corresponds to a GAS initial subspace. }

When we choose a basis according to the DID, the matrices $M_k$ of the OSR acquire a specific block structure, in which one block depends on which case has stopped the construction. 

If the DID terminates successfully, as in case two, then we get:
{\color{black} \footnotesize
\begin{align}
M_k&=\left[\begin{array}{cccccc}
M_{k,S} & M_{k,P_1} & 0 & \dots & \dots & 0 \\
0 & M_{k,T_1} & M_{k,P_2} & 0 & \dots & 0 \\
0 & M_{k,Q_{2,1}} & M_{k,T_2} & \ddots & & \vdots \\
\vdots & \vdots &   & & \ddots &  \\
0 & M_{k,Q_{N-1,1}} & M_{k,Q_{N-1,2}} & \dots & M_{k,T_{N-1}} & M_{k,P_{N}} \\
0 & M_{k,Q_{N,1}} & M_{k,Q_{N,2}} & \dots & M_{k,Q_{N,N-1}} & M_{k,T_{N}}
\end{array}\right] \nonumber\\
k&=1,\ldots K, \label{eq:succDIDmat}
\end{align}}
with $\cap_k \ker(M_{k,P_i}) = 0 $ for $i=1,\ldots N$. {\ftr The zero blocks in the first column indicate that $\Hi_S$ is invariant, while those over the first upper diagonal show that a state with support on one of the $\Hi_{T_i}$ can only transition to the preceding subspace in the chain $\Hi_{T_{i-1}},$ and cannot e.g. ``jump'' directly to $\Hi_S$.}

If instead the decomposition is terminated due to case one, we obtain:
{\color{black} \footnotesize
\begin{align}
M_k&=\left[\begin{array}{cccccc}
M_{k,S} & M_{k,P_1} & 0 & \dots & \dots & 0 \\
0 & M_{k,T_1} & M_{k,P_2} & 0 & \dots & 0 \\
0 & M_{k,Q_{2,1}} & M_{k,T_2} & \ddots & & \vdots \\
\vdots & \vdots &   & & \ddots &  \\
0 & M_{k,Q_{N-1,1}} & M_{k,Q_{N-1,2}} & \dots & M_{k,T_{N-1}} & 0 \\
0 & M_{k,Q_{N,1}} & M_{k,Q_{N,2}} & \dots & M_{k,Q_{N,N-1}} & M_{k,T_{N}}
\end{array}\right] \nonumber\\
k&=1,\ldots K.\label{eq:failDIDmat}
\end{align}}
In this case, the first $N-1$ zero blocks in the last column, on top of $M_{k,T_{N}},$ for {\em all $k$}, imply the invariance of the last subspace, $\Hi_{T_N}$.

\subsection{DID Properties}\label{sec:DIDprop}

As in the previous case, the obtained decomposition allows us to decide the asymptotic stability of the face.
\begin{proposition}\label{prop:didgas}
$\Hi_S$ is GAS if and only if the DID is terminated successfully.
\end{proposition}
\proof
If the DID is terminated due to case 1 the matrices have the structure \eqref{eq:failDIDmat}. By a reordering of the basis, and hence of the blocks, and by using Proposition \ref{prop:genInva}, is easily seen that $\Hi_{T_N}$ is invariant. Hence $\Hi_S$ cannot be GAS.

If instead the DID is terminated successfully it is possible to verify that no fixed point has support in $\mathcal{H}_R$, and by the results of \citep{bolognani-arxiv} this is equivalent to show that the target subspace is attractive. Suppose by contradiction that $\rho$ is a density operator with support in $\Hi_R$ and a fixed point for the map, i.e.:
\beq
\supp(\rho)\subset\bigoplus\limits_{i=2}^N\Hi_{T_i}.
\eeq
Then there exists a maximal $j$ between $2$ and $N$ (note that the next subspace is minimal) such that
\beq \label{eq:suppDID}
\supp(\rho)\subset\bigoplus\limits_{i=j}^N\Hi_{T_i}.
\eeq
Then $\rho_{T_j}$ (the block of $\rho$ corresponding to $\Hi_{T_j},$ when it is decomposed accordingly to the DID) is not zero, otherwise, due to positiveness of $\rho$, the corresponding columns and rows should be zero and \eqref{eq:suppDID} would hold with $j+1$. Using the block structure \eqref{eq:succDIDmat} and the fact that $\rho$ is a fixed point, we must have:
\beq
T(\rho)_{T_{j-1}}=\sum\limits_k M_{k,P_j}\rho_{T_j}M_{k,P_j}^\dagger=0,
\eeq
This is a sum of PSD matrices so any terms must be zero.\\
If $\rho_{T_j}=C C^\dagger$ with $C\in \mathfrak{B}(\mathcal{H}_{T_j})$ using the last
\beq
M_{k,P_j}C = 0,
\eeq
for any $k$, then the columns of $C$ should be in $\cap_k \ker(M_{k,P_i})$ which is impossible by construction, since at each step we choose $\cap_k \ker(M_{k,P_i}) = \{0\}.$
\qed
Suppose now that the DID is completed successfully, and that we are interested in estimating the convergence speed to the target subspace. Fix a state $\rho$, with support {\color{black} in a fixed $\Hi_{T_i}$, i.e that verify:
\begin{equation}\label{eq:sand}
\Pi_i\rho\Pi_i=\rho,
\end{equation}
if $\Pi_i$ is the orthogonal projection onto $\Hi_{T_i}$.} Consider the probability of finding the system in the preeceding subspace $\Hi_{T_{i-1}}$ after one application of $T$: 
\begin{equation}\label{eq:probDID}
{\mathbb P}_{T(\rho)}(\Pi_{T_{i-1}})=\tr(\Pi_{T_{i-1}}T(\rho)),
\end{equation}
where $\Pi_{T_{i-1}}$ is the orthogonal projection onto $\Hi_{T_{i-1}}$. It is possible to obtain bounds on the growth of this probability (recall it was zero before the action of the map):
\begin{align}
{\mathbb P}_{T(\rho)}(\Pi_{T_{i-1}})&=\tr(\Pi_{T_{i-1}}T(\rho)) \nonumber \\
 &=\tr(\sum\limits_{k}M_{k,P_{i}}\rho_{T_{i}}M_{k,P_{i}}^\dagger) \nonumber \\
 &=\tr(\sum\limits_{k}M_{k,P_{i}}^\dagger M_{k,P_{i}}\rho_{T_{i}})
\end{align}

{\color{black} In fact, the increment in probability is at least 
\begin{equation}
\gamma_{i} \tr(\rho_{T_i}),
\end{equation}
if $\gamma_i$ is the least eigenvalue of $\sum_{k}M_{k,P_{i}}^\dagger M_{k,P_{i}}$,} as is seen putting $\sum_{k}M_{k,P_{i}}^\dagger M_{k,P_{i}}$ in its diagonal form. In the same way  it is possible to give an upper bound for \eqref{eq:probDID} by the maximum eigenvalue of $\sum_{k}M_{k,P_{i}}^\dagger M_{k,P_{i}}$. Note that these bounds are always meaningful: $\sum_{k}M_{k,P_{i}}^\dagger M_{k,P_{i}}$ is positive definite by construction, so its lowest eigenvalue cannot be zero. Moreover, exists a density operator (e.g. the projection on the subspace generated by an eigenvector corresponding to $\gamma_{i}$) for which the lower bound is reached, and the same is true for the upper bound (this shows also that the maximum eigenvalue is less than or equal to one). 

In the light of these observations, the $\gamma_i$ are indications of the (maximal and minimal) probabilities that a transition from $\Hi_{T_i}$ to $\Hi_{T_{i-1}}$ occurs. Knowing all these {\ic {\em transition rates}} we can use the minimal ones to find the convergence bottlenecks, and estimate the worst-case time needed to reach $\Hi_S$ starting from any state. 

{\color{black} It should be {\ic noted that} the bounds derived are significant only when \eqref{eq:sand} holds, due to the possible presence of blocks which connect the subspaces $\Hi_{T_i}$ in the opposite way, namely making probability flow down the chain. However, since we assume $\Hi_S$ to be GAS, the transitions towards it ``dominates'' the dynamics, so the $\gamma_i$ can be effectively used to estimate the convergence speed.}
\subsection{Dual Characterization}\label{sec:dualDID}
The DID can be also studied, and in fact characterized, in the Heisenberg picture. In this dual framework some properties become more explicit, e.g. its independence from the chosen OSR. The following characterization of invariant subspaces which refers to the dual map will be needed later.
\begin{proposition}\label{prop:InvSupp}
Let $T$ be a CPTP map, and $T^*$ its dual. A subspace $\Hi_S$ is invariant for $T$ if and only if for any $n$
\beq
T^{*n+1}(\Pi_S)\geq T^{*n}(\Pi_S).
\eeq
\end{proposition}
\proof
By unitality of the dual map we have
\begin{align}
\begin{split}\label{eq:unit}
\Pi_S+\Pi_R&=I=T^*(I)=T^*(\Pi_S+\Pi_R)\\&=T^*(\Pi_S)+T^*(\Pi_R).
\end{split}
\end{align}
Given Proposition \ref{prop:genInva}, it is easy to see that if $\Hi_S$ is invariant for $T$ then $\Hi_R$ is invariant for $T^*.$ Hence we have that $T^*(\Pi_R)=T_R^*(\Pi_R),$ which has support only on $\Hi_R.$ Hence, for \eqref{eq:unit} to be true, it must be
\beq
T^*(\Pi_{S})=T_S^*(\Pi_S) + T_{SR}^*(\Pi_S)=\Pi_S + T_{SR}^*(\Pi_S).
\eeq
Applying $n$-times $T^*$ and using the invariance of $\Hi_R$ for the dual map, we thus have:
\beq
T^{*n}(\Pi_{S})=\Pi_S + \sum\limits_{i=1}^{n-1}T^{*i}_R(T_{SR}^*(\Pi_S)).
\eeq
Since $T^{*n}_R(T_{SR}^*(\Pi_S)) \ge 0$ for any $n$, the sequence $T^{*n}(\Pi_S)$ is non decreasing.\\
Suppose now
\beq \label{eq:theHyp}
T^*(\Pi_S) \geq \Pi_S.
\eeq
Let $\Hi_R$ be the orthogonal complement of  $\Hi_S$ so that
\beq
\Pi_S+\Pi_R=I,
\eeq
and then
\beq \label{eq:theTrick}
T^*(\Pi_S)+T^*(\Pi_R)=I.
\eeq
Rewriting these two terms in their block form
\beq
T^*(\Pi_S)=\sum\limits_k\left[\begin{array}{cc}
M_{k,S}^\dagger M_{k,S} & M_{k,S}^\dagger M_{k,P} \\
M_{k,P}^\dagger M_{k,S} & M_{k,P}^\dagger M_{k,P}
\end{array}\right],
\eeq
\beq \label{eq:dec2}
T^*(\Pi_R)=\sum\limits_k\left[\begin{array}{cc}
M_{k,Q}^\dagger M_{k,Q} & M_{k,Q}^\dagger M_{k,R} \\
M_{k,R}^\dagger M_{k,Q} & M_{k,R}^\dagger M_{k,R}
\end{array}\right],
\eeq
Now by unitality
\beq
\sum\limits_k M_{k,S}^\dagger M_{k,S} \leq I,
\eeq
and from \eqref{eq:theHyp}
\beq
\sum\limits_k M_{k,S}^\dagger M_{k,S} \geq I,
\eeq
 so 
\beq
\sum\limits_k M_{k,S}^\dagger M_{k,S} = I.
\eeq
But now \eqref{eq:theTrick}, together with \eqref{eq:dec2}, implies
\beq
M_{k,Q}^\dagger M_{k,Q} = 0,
\end{equation}
for any $k$, and by Proposition \ref{prop:genInva} invariance of ${\cal H}_S$ is proved.
\qed
So a subspace is invariant if the sequence generated by the corresponding orthogonal projection is monotonically non decreasing.
From this result immediately follows:
\begin{cor}\label{cor:theCorollary}
Let $T$ be a CPTP map and $\Hi_S$ a subspace:
\begin{enumerate}[i)]
\item $\Hi_S$ is invariant if and only if $T^*(\Pi_S) \geq \Pi_S$.
\item If $\Hi_S$ is invariant then for any $n$
\beq
\label{eq:theCorInc}
\supp(T^{*n}(\Pi_S)) \subset \supp(T^{*n+1}(\Pi_S)).
\eeq
\end{enumerate}
\end{cor}
The first point is already known, and projections satisfying this property are called \emph{sub-harmonic} \cite{wolf-notes}.
Using these results, we can give a dual characterization of the DID.
\begin{proposition}
Let $T$ be a CPTP map, $\Hi_S$ a GAS subspace and consider the decomposition induced by the DID. Then for $n\leq N$
\begin{equation}\label{eq:didDualProp}
\supp(T^{*n}(\Pi_{S}))=\Hi_S\oplus\bigoplus\limits_{i=1}^n \Hi_{T_i}
\end{equation} 
and for $n>N$ the support is the whole $\Hi$. \end{proposition}
\proof
We shall prove that \eqref{eq:didDualProp} holds by induction on $n$.\\
First consider the case $n=1$. By using the matrix block-decomposition provided in \eqref{eq:succDIDmat}, it is easy to show that:
{\color{black} \begin{equation}\label{eq:firstStep}
\begin{gathered}
T^*(\Pi_S)=\sum\limits_k M_k^{\dagger} \Pi_S M_k \\
=\sum\limits_k \left[ \begin{array}{ccccc}
M_{k,S}^\dagger M_{k,S}   & M_{k,S}^\dagger M_{k,P_1}   & 0   		& \cdots & 0 \\
M_{k,P_1}^\dagger M_{k,S} & M_{k,P_1}^\dagger M_{k,P_1} & 0 		& \cdots & 0 \\
0                         &          0                  & 0 		& \cdots & 0 \\
\vdots                    &          \vdots             & \ddots 	& 		 & \vdots \\
0						  & 		0					& \cdots    &        &   0
\end{array} \right].
\end{gathered}
\end{equation} }%
Recall that by \eqref{eq:theCorInc}
\begin{equation}\label{eq:firstInc}
\mathcal{H}_S \subset \supp(T^*(\Pi_S)),
\end{equation}
so, by \eqref{eq:firstStep}, it suffices to show that $\mathcal{H}_{T_1}$ is contained in $\supp(T^*(\Pi_S))$. Choosing a set of vector $ |\varphi_{1,h}\rangle$ in $\mathcal{H}_{T_1}$, applying $T^*(\Pi_S)$ results in:
\begin{equation}
T^*(\Pi_S)|\varphi_{1,h}\rangle = \left[ \begin{array}{c}
M_{k,S}^\dagger M_{k,P_1}|\varphi_{1,h}\rangle \\
M_{k,P_1}^\dagger M_{k,P_1}|\varphi_{1,h}\rangle \\
0\\
\vdots\\
0
\end{array}\right]=\left[\begin{array}{c}
|\psi_{1,h} \rangle \\
|\phi_{1,h} \rangle \\
0\\
\vdots\\
0
\end{array}\right]
\end{equation}
By a proper choice of $ |\varphi_{1,h}\rangle$, it is possible to obtain that the $|\phi_{1,h} \rangle$ are a basis for $\mathcal{H}_{T_1},$ since by construction the range of $\sum_k M_{k,P_1}^\dagger M_{k,P_1}$ is $\mathcal{H}_{T_1}$. Hence, the case $n=1$ is completed.\\ 
Now assume \eqref{eq:didDualProp} true for $n<l$, and call $\Pi_{i}$ the projection on $\mathcal{H_S}\oplus\bigoplus_{j=1}^i\mathcal{H}_{T_j}$ so that
\begin{equation}
\supp(T^{*l-1}(\Pi_{S}))=\mathcal{H_S}\oplus\left(\bigoplus_{j=1}^{l-1}\mathcal{H}_{T_j}\right)=\supp(\Pi_{l-1}),
\end{equation}
and for some real constant $c>0$ and $C>0$
\begin{equation}
c\Pi_{l-1}\leq T^{*l-1}(\Pi_{S}) \leq  C\Pi_{l-1}.
\end{equation}
This implies that the support of $T^*(\Pi_{l-1})$ and $T^{*l}(\Pi_{S})$ is the same and the statement is proved if  $\supp(T^*(\Pi_{l-1}))$ coincides with $\mathcal{H_S}\oplus\bigoplus_{j=1}^{l}\mathcal{H}_{T_j}$, which can be verified proceeding as above.
By \eqref{eq:theCorInc} in Corollary \ref{cor:theCorollary}, we have:
\begin{equation}\label{eq:inclusionProj}
\mathcal{H_S}\oplus\bigoplus_{j=1}^{l-1}\mathcal{H}_{T_j} \subset \supp(T^{*l}(\Pi_{S})) = \supp(T^*(\Pi_{l-1})).
\end{equation}
Applying $T^*(\Pi_{l-1})$ to a set of vector $ |\varphi_{l,h}\rangle$ in $\mathcal{H}_{T_l}$ results in:
{\color{black}
\begin{align}
T^*(\Pi_{l-1})|\varphi_{l,h}\rangle &=\sum\limits_k \left[ \begin{array}{c}
0 \\
M_{k,Q_{l-1,1}}^\dagger M_{k,P_l}|\varphi_{l,h}\rangle  \\
M_{k,Q_{l-1,1}}^\dagger M_{k,P_l}|\varphi_{l,h}\rangle  \\
\vdots \\
M_{k,T_{l-1}}^\dagger M_{k,P_l}|\varphi_{l,h}\rangle  \\
M_{k,P_l}^\dagger M_{k,P_l}|\varphi_{l,h}\rangle \\
0\\
\vdots\\
0
\end{array}\right] \nonumber \\
&=
\left[ \begin{array}{c}
|\psi_{l,h} \rangle \\
|\phi_{l,h} \rangle \\
0\\
\vdots\\
0
\end{array} \right],\label{eq:latter}
\end{align}}%
where $|\phi_{l,h} \rangle=\sum_k M_{k,P_l}^\dagger M_{k,P_l}|\varphi_{l,h}\rangle$ and $|\psi_{l,h} \rangle $ accounts for the first blocks of elements. By the \eqref{eq:latter} ${|\psi_{l,h} \rangle \oplus |\phi_{l,h} \rangle}$ are in $\supp(T^*(\Pi_{l-1}))$. Again choosing properly $|\varphi_{l,h}\rangle$, and noting that $\sum\limits_k M_{k,P_l}^\dagger M_{k,P_l}$ is strictly positive, we prove that the desired property \eqref{eq:didDualProp} holds for $l$ as well. Finally, the last statement follows directly from the first and the hypothesis of $\mathcal{H}_S$ being GAS, which in turn implies that the DID algorithm runs to completion and $\mathcal{H}_S \bigoplus\limits_{i=1}^N \mathcal{H}_{T_i}=\mathcal{H}$.
\qed
So the DID is determined by, and is in fact equivalent to, the sequence of supports \eqref{eq:didDualProp}. Since it depends only from the form of $T^*$ we readily obtain that the DID is independent from the chosen OSR. From the last result another useful property is also obtained.
\begin{cor}\label{cor:lastDID}
Suppose $\mathcal{H}_S$ is an invariant subspace for the CPTP map $T$. Then it is GAS if and only if the sequence $\supp(T^{*n}(\Pi_S))$ is strictly increasing until it covers the whole space.
\end{cor}
\proof
One implication is just a restatement of the last proposition.
For the converse, it suffices to follow the proof, taking into account that the failure of the DID returns the block structure of the form \eqref{eq:failDIDmat}, from which the sequence $\supp(T^{*n}(\Pi_S))$ cannot cover the whole space.
\qed

\section{Asymptotic Probabilities}
\label{sec:asProb}
The two decompositions we introduced in the previous Sections essentially study the transient structure and dynamical behavior of a subspace $\Hi_{R}$, complementary to $\Hi_S$, when the last one is invariant. When $\Hi_S$ is GAS, we can further investigate its ``internal'' asymptotic behavior, and understand where the system will converge to, and with which probability.


If a subspace $\Hi_S$ is GAS, it must contain the supports of all fixed points. It is well known (see e.g. \citep{wolf-notes,viola-IPSlong}), that for a CPTP map $T$ the subspace of fixed points has a structure of the form:
\beq
\mathcal{F}_T = U\left(\bigoplus\limits_{k=1}^K \mathcal{M}_{d_k}\otimes\rho_k\oplus 0\right)U^\dagger,
\eeq
where $U$ is a unitary operator, $\mathcal{M}_{d_k}$ stands for the full algebra of complex matrices on $\mathbb{C}^{d_k}$ and $\rho_k$ are positive definite density matrices. When the action of $T$ is restricted to the convex, invariant set of density operators, to find the invariant sets it is enough to substitute the full algebras with the sets of density operators contained in the $\mathcal{M}_{d_k}$.
This highlights a direct-sum structure for the {\em minimal} GAS subspace (or {\em collecting} subspace \cite{baumgartner-1}):
\beq\label{eq:hildecasym}
\Hi_S=\bigoplus\limits_{k=1}^K\Hi_{S_i},
\eeq
with all the $\Hi_{S_i}$ invariant,  being supports of invariant states. 
{\ftr This decompostion of the minimal GAS subspace is of particular interest in quantum information applications, since to each component $\Hi_{S_i}$ remains associated an information preserving structure \cite{viola-IPS}: namely, the subspace $\Hi_{S_i}$ contains a perfectly noiseless subsystem of dimension $d_k$ that allows to store and preserve quantum information \cite{viola-generalnoise}. 

An interesting question is whether the dynamics will drive the state into a protected subsystem of interest, and with what probability this will happen, depending on the initial state of the dynamics. The first part of the question can be addressed by checking if the support of the noiseless subsystem is GAS. If this is not case, it is possible} to derive explicit formulas for the asymptotic probabilities to find the state in one of the $\Hi_{S_i}$, i.e. to evaluate
\beq
\lim\limits_{n\to \infty} \Tr(\Pi_{S_i}T^n(\rho))
\eeq
given the initial state $\rho$, where $\Pi_{S_i}$ is the orthogonal projection on $\Hi_{S_i}$.  
{\ftr A particular case of interest emerges when all the subspaces are one-dimensional, and the evolution induces {\em decoherence} with respect to this (partial) orthogonal preferred basis. Given an initial state, what is the probability of finding it in one of the orthogonal and pure pointer states?

In deriving a suitable tool to answer these questions, a} key preliminary result is represented by the following Lemma.

\begin{lemma}
Let $T$ be a CPTP map, $\Hi=\Hi_S\oplus \Hi_R,$ where $\Hi_S=\bigoplus\limits_{k=1}^K\Hi_{S_i}$ {\ftr with each} of the  $\Hi_{S_i}$ being invariant. If $\Pi_{S_i}$ is the orthogonal projection on $\Hi_{S_i}$ then for any $i$ 
\beq
T^*(\Pi_{S_i})=\Pi_{S_i}+T^*_{SR}(\Pi_{S_i}),
\eeq
\end{lemma}
\proof
We will first explicitly prove the statement for $K=2.$ Given that the $\Hi_{S_{1,2}}$ are invariant, with respect to the orthogonal sum $\Hi_S=\bigoplus\limits_{k=1}^K\Hi_{S_i}$ the matrices $M_k$ have the block-structure:
\beq \label{eq:twoIn}
\left[ \begin{array}{ccc}
M_{k,S_1}  & 0 & M_{k,P_1} \\
0 & M_{k,S_2} & M_{k,P_2} \\
0 & 0 & M_{k,R}
\end{array}
\right].
\end{equation}
{\color{black} Taking into account the unitality condition, by the block form we derive the relations:
\begin{align}
\sum_k M_{k,S_1}^\dagger M_{k,S_1}=I,\\
\sum_k M_{k,P_1}^\dagger M_{k,S_1} = 0, \\
\sum_k M_{k,S_2}^\dagger M_{k,S_2}=I,\\
\sum_k M_{k,P_2}^\dagger M_{k,S_2} = 0,
\end{align} }
Let us focus on $\Hi_{S_1},$ as the same reasoning applies to $\Hi_{S_2}$ up to a relabeling. In the same block-representation, the projection of interest is
\beq
\begin{aligned}
\Pi_{S_1}=\left[\begin{array}{ccc}
I & 0 & 0 \\
0 & 0 & 0 \\
0 & 0 & 0
\end{array}\right],
\end{aligned}
\eeq
{\color{black} we thus have:
\begin{align}
T^*(\Pi_1)&=
\sum\limits_k
\left[ \begin{array}{ccc}
M_{k,S_1}^\dagger M_{k,S_1}  & 0 & M_{k,S_1}^\dagger M_{k,P_1} \\
0 & 0 & 0 \\
M_{k,P_1}^\dagger M_{k,S_1} & 0 & 
M_{k,P_1}^\dagger M_{k,P_1}
\end{array} \right] \nonumber \\
&=\Pi_{S_1}+T^*_{SR}(\Pi_{S_1}).
\end{align}}
In the general case $\Hi_S = \bigoplus_i \Hi_{S_i} $, for any ${j=1, \ldots ,K}$  we can consider the decomposition ${\Hi_S=\Hi_{S_j}\oplus\bigoplus_{i\neq j}\Hi_{S_i}}$. These two orthogonal subspaces in the sum are both invariant so by the reasoning above the evolution of $\Pi_{S_j}$ has the desired form. \qed

 Using the above Lemma, we can then provide a formula to compute the asymptotic probability analitically, depending on the initial state.

\begin{proposition}\label{prop:asymProb}
Under the hypothesis of the preceding lemma, assume also that $\Hi_S$ is GAS, then 
{\color{black}
\begin{equation}\label{eq:limRes}
\begin{gathered}
\lim\limits_{n\to \infty} \Tr(\Pi_{S_i}T^n(\rho))\\=\Tr(\Pi_{S_i}\rho_S)+\Tr(\Pi_{S_i}T_{SR}((I-T_R)^{-1}(\rho_R))).
\end{gathered}
\end{equation} }
\end{proposition}
\proof
The limit of $\Pi_{S_i}$ under the action of $T^*$ is easily computed:
\beq
\begin{aligned}
T^*(\Pi_{S_i})&=\Pi_{S_i} + T_{SR}^*(\Pi_{S_i}),\\
T^{*2}(\Pi_{S_i})&=\Pi_{S_i} + (T^*_R(T_{SR}^*(\Pi_{S_i}))+T_{SR}^*(\Pi_{S_i})),\\
&\vdots\\
T^{*n}(\Pi_{S_i})&=\Pi_{S_i} + \sum\limits_{k=0}^{n-1} T^{*k}_R(T_{SR}^*(\Pi_{S_i})).
\end{aligned}
\eeq
Letting $n$ go to infinity
{\color{black} \begin{align}
\lim\limits_{n \to \infty} T^{*n}(\Pi_{S_i}) &= \lim\limits_{n \to \infty}\Pi_{S_i}+(\sum\limits_{k=0}^{n-1} T^{*k}_R(T_{SR}^*(\Pi_{S_i})))\nonumber \\ 
&=\Pi_{S_i}+((I-T^{*}_R)^{-1}(T_{SR}^*(\Pi_{S_i}))),
\end{align} }
where the last equality follows from the fact that since $\sigma(T_R)<1$
\beq
\sum\limits_{k=0}^{\infty} T^{*k}_R=(I-T^{*}_R)^{-1}.
\eeq
By using the relation:
\beq
\lim\limits_{n \to \infty}\tr(\Pi_{S_i}T^n(\rho))=\lim\limits_{n \to \infty}\tr(T^{*n}(\Pi_{S_i})\rho),
\eeq
we get the statement.
\qed
 In conclusion, the asymptotic probability of converging to an invariant subspace inside the the minimal GAS subspace is given by the sum of two terms: the initial probability of finding the state there (the term $\Tr(\Pi_{S_i})$), plus a (linear) term that can be computed explicitly knowing the map decomposition as in \eqref{eq:TrDef}-\eqref{eq:TsrDef}.

\section{An Illustrative Example}\label{sec:example}

In this section we put our results at work, showing how they can be employed to study the dynamical behavior of different faces of the positive cone and their asymptotic probabilities.
\subsection{Description of the dynamics}
Consider a 7 level quantum system associated to the Hilbert space $\Hi = \spn(\{ \ket{j} \}_{j=1}^{7})$, on which, within each fixed time step, one of the following ``noise actions'' may occur:
\begin{enumerate}[i)]
\item with probability $\gamma_1 < 1,$ level 1, 3 and 2, 4 are swapped,
\item with probability $\gamma_2 < 1,$ level 3 decays to 1 and 4 to 2,
\item with probability $\gamma_3 \ll 1,$ level 5 decays to level 4 and 3 in the same proportion,
\item with probability $\gamma_4 < 1,$ level 6 decays to 5,
\item with probability $\gamma_5 < 1,$ level 7 decays to 5;
\end{enumerate}
where $\sum_i \gamma_i = 1$, $\gamma_i > 0$ for any $i$ and $\gamma_3 < \gamma_4 <\gamma_5$.
An OSR for the map $T$ jointly describing these processes can be obtained by the following matrices, associated to each of the processes in the ordered basis for $\Hi$ given above (see e.g. \cite{nielsen-chuang}, Chapter 8 for details on phenomenological description of noise actions):
\begin{enumerate}[i)]
\item $N_1=
\left[ \begin{array}{ccccccc}
0 & 0 & 1 & 0 & 0 & 0 & 0\\
0 & 0 & 0 & 1 & 0 & 0 & 0\\
1 & 0 & 0 & 0 & 0 & 0 & 0\\
0 & 1 & 0 & 0 & 0 & 0 & 0\\
0 & 0 & 0 & 0 & 1 & 0 & 0\\
0 & 0 & 0 & 0 & 0 & 1 & 0\\
0 & 0 & 0 & 0 & 0 & 0 & 1
\end{array}\right]$;
\item $N_2=
\left[ \begin{array}{ccccccc}
0 & 0 & 1 & 0 & 0 & 0 & 0\\
0 & 0 & 0 & 1 & 0 & 0 & 0\\
0 & 0 & 0 & 0 & 0 & 0 & 0\\
0 & 0 & 0 & 0 & 0 & 0 & 0\\
0 & 0 & 0 & 0 & \frac{1}{\sqrt{2}} & 0 & 0\\
0 & 0 & 0 & 0 & 0 & \frac{1}{\sqrt{2}} & 0\\
0 & 0 & 0 & 0 & 0 & 0 & \frac{1}{\sqrt{2}}
\end{array}\right]$,\\ $N_3=
\left[ \begin{array}{ccccccc}
1 & 0 & 0 & 0 & 0 & 0 & 0\\
0 & 1 & 0 & 0 & 0 & 0 & 0\\
0 & 0 & 0 & 0 & 0 & 0 & 0\\
0 & 0 & 0 & 0 & 0 & 0 & 0\\
0 & 0 & 0 & 0 & \frac{1}{\sqrt{2}} & 0 & 0\\
0 & 0 & 0 & 0 & 0 & \frac{1}{\sqrt{2}} & 0\\
0 & 0 & 0 & 0 & 0 & 0 & \frac{1}{\sqrt{2}}

\end{array}\right]$;
\item $N_4=
\left[ \begin{array}{ccccccc}
0 & 0 & 0 & 0 & 0 & 0 & 0\\
0 & 0 & 0 & 0 & 0 & 0 & 0\\
0 & 0 & 0 & 0 & \frac{1}{\sqrt{2}} & 0 & 0\\
0 & 0 & 0 & 0 & \frac{1}{\sqrt{2}} & 0 & 0\\
0 & 0 & 0 & 0 & 0 & 0 & 0\\
0 & 0 & 0 & 0 & 0 & 0 & 0\\
0 & 0 & 0 & 0 & 0 & 0 & 0
\end{array}\right]$,\\ $N_5=
\left[ \begin{array}{ccccccc}
1 & 0 & 0 & 0 & 0 & 0 & 0\\
0 & 1 & 0 & 0 & 0 & 0 & 0\\
0 & 0 & 1 & 0 & 0 & 0 & 0\\
0 & 0 & 0 & 1 & 0 & 0 & 0\\
0 & 0 & 0 & 0 & 0 & 0 & 0\\
0 & 0 & 0 & 0 & 0 & 1 & 0\\
0 & 0 & 0 & 0 & 0 & 0 & 1\\
\end{array}\right]$;
\item $N_6=
\left[\begin{array}{ccccccc}
0 & 0 & 0 & 0 & 0 & 0 & 0\\
0 & 0 & 0 & 0 & 0 & 0 & 0\\
0 & 0 & 0 & 0 & 0 & 0 & 0\\
0 & 0 & 0 & 0 & 0 & 0 & 0\\
0 & 0 & 0 & 0 & 0 & 1 & 0\\
0 & 0 & 0 & 0 & 0 & 0 & 0\\
0 & 0 & 0 & 0 & 0 & 0 & 0\\
\end{array}\right]$,\\ $N_7=
\left[\begin{array}{ccccccc}
1 & 0 & 0 & 0 & 0 & 0 & 0\\
0 & 1 & 0 & 0 & 0 & 0 & 0\\
0 & 0 & 1 & 0 & 0 & 0 & 0\\
0 & 0 & 0 & 1 & 0 & 0 & 0\\
0 & 0 & 0 & 0 & 1 & 0 & 0\\
0 & 0 & 0 & 0 & 0 & 0 & 0\\
0 & 0 & 0 & 0 & 0 & 0 & 1\\
\end{array}\right]$;
\item $N_8=
\left[\begin{array}{ccccccc}
0 & 0 & 0 & 0 & 0 & 0 & 0\\
0 & 0 & 0 & 0 & 0 & 0 & 0\\
0 & 0 & 0 & 0 & 0 & 0 & 0\\
0 & 0 & 0 & 0 & 0 & 0 & 0\\
0 & 0 & 0 & 0 & 0 & 0 & 1\\
0 & 0 & 0 & 0 & 0 & 0 & 0\\
0 & 0 & 0 & 0 & 0 & 0 & 0\\
\end{array}\right]$,\\ $N_9=
\left[\begin{array}{ccccccc}
1 & 0 & 0 & 0 & 0 & 0 & 0\\
0 & 1 & 0 & 0 & 0 & 0 & 0\\
0 & 0 & 1 & 0 & 0 & 0 & 0\\
0 & 0 & 0 & 1 & 0 & 0 & 0\\
0 & 0 & 0 & 0 & 1 & 0 & 0\\
0 & 0 & 0 & 0 & 0 & 1 & 0\\
0 & 0 & 0 & 0 & 0 & 0 & 0\\
\end{array}\right]$.
\end{enumerate}
Defining the probability-weighed matrices $M_1 = \sqrt{\gamma_1} N_1$, $M_2=\sqrt{\gamma_2} N_2$, $M_3=\sqrt{\gamma_2} N_3$, $M_4=\sqrt{\gamma_3} N_4$, $M_5=\sqrt{\gamma_3} N_5$, $M_6=\sqrt{\gamma_4}N_6$, $M_7=\sqrt{\gamma_4}N_7$, $M_8=\sqrt{\gamma_5}N_8$ and $M_9=\sqrt{\gamma_5}N_9$ we obtain a representation for the whole process $T$.

\subsection{Checking GAS}

By looking at the structure of the matrices, it is easy to note that the subspace $\Hi_{S_1}=\spn(\{\ket{1},\ket{3}\})$ is invariant. This allows us to employ the results of section \ref{sec:dualDID} in order to to check if it is also GAS. We must look at the sequence of supports $T^{*n}(\ket{1}\bra{1}+\ket{3}\bra{3})$. We obtain:
\beqan
\supp(T^*(\ket{1}\bra{1}+\ket{3}\bra{3}))&=&\spn(\ket{1},\ket{3},\ket{5}),\\
\supp(T^{*2}(\ket{1}\bra{1}+\ket{3}\bra{3}))&=&\spn(\ket{1},\ket{3},\ket{5},\ket{6},\ket{7}),\\
\supp(T^{*3}(\ket{1}\bra{1}+\ket{3}\bra{3}))&=&\spn(\ket{1},\ket{3},\ket{5},\ket{6},\ket{7}).
\eeqan
Since this sequence stops before covering the whole $\Hi$, by Corollary \ref{cor:lastDID} $\Hi_{S_1}$ is not GAS.

\subsection{Nested Faces}
It is interesting find out what is the minimal subspace that contains $\Hi_{S_1}$ and is GAS. This can be done using the nested faces construction, thanks to the results in section \ref{sec:NFprop}. Decomposition \eqref{eq:NFdec} returns in this case to the following subspaces, each characterized by the spectral radius of the corresponding $T_{R_i}$:
\beqan
\Hi_{T_1}=&\spn(\{\ket{2},\ket{4}\}) \,\,\,\, &\sigma(T_{R_1}) = 1, \\
\Hi_{T_2}=&\spn(\{\ket{5}\}) \,\,\,\, &\sigma(T_{R_2}) = 1-\gamma_3, \\
\Hi_{T_3}=&\spn(\{\ket{6}\}) \,\,\,\, &\sigma(T_{R_3}) = 1-\gamma_4, \\
\Hi_{T_4}=&\spn(\{\ket{7}\}) \,\,\,\, &\sigma(T_{R_4}) = 1-\gamma_5. \\
\eeqan
As expected, given Proposition \ref{prop:leastInvNF}, $\sigma(T_{R_1}) = 1$; moreover, the same proposition permits to obtain the minimal GAS subspace, which is $\Hi_S=\Hi_{S_1}\oplus\Hi_{S_2}$.  In our case where $\Hi_{S_2}=\Hi_{T_1}=\spn(\{\ket{2},\ket{4}\}),$ we obtain:
\[\Hi_{S}=\spn(\{\ket{1},\ket{3},\ket{2},\ket{4}\}).\]

The subspace $\Hi_S$ can be used as the starting point for the DID; doing so, decomposition \eqref{eq:DIDdec} is given by:
\beqan
\Hi_{T_1'}&=&\spn(\{\ket{5}\}),\\
\Hi_{T_2'}&=&\spn(\{\ket{6},\ket{7}\}).
\eeqan
For any of these subspaces there is a minimal and a maximal transition {\ic rate}, as explained in section \ref{sec:DIDprop}, the least of which has value $\gamma_3$ (in this case it can be read out directly from the form of the dynamics, and in particular $M_4$). A comparison with the maximal spectral radius of the nested faces decomposition shows that both the constructions give the same estimation for the covergence speed towards $\Hi_S$.

\subsection{Asymptotic probabilities}
Knowing that $\Hi_S$ is GAS, it is possible to use the results in section \ref{sec:asProb}, to evaluate the asymptotic probabilities of the two subspaces $\Hi_{S_1}$ and $\Hi_{S_2}$. In order to make the structure of the fixed-point set explicit, it is useful to note that representing the dynamics restricted to $\Hi_S$ in the basis $ \{ \ket{1},\ket{3},\ket{2},\ket{4} \}$, one directly obtains a tensor structure. In fact, by relabeling these four states as $\ket{1}=\ket{0_N}\otimes\ket{0_F}$, $\ket{2}=\ket{1_N}\otimes\ket{0_F}$, $\ket{3}=\ket{0_N}\otimes\ket{1_F}$ and $\ket{4}=\ket{1_N}\otimes\ket{1_F}$, results in a decomposition of $\Hi_S$ in two ``virtual'' subsystem of dimension 2: $\Hi_S=\Hi_N\otimes\Hi_F$. With respect to this decomposition the matrices that generates the dynamics {\em inside} $\Hi_S$ can be written as:
\beqan
B_1=\sqrt{\gamma_1}I_2\otimes\left[\begin{array}{cc}
0 & 1\\ 
1 & 0\end{array}\right],\\
B_2=\sqrt{\gamma_2}I_2\otimes\left[\begin{array}{cc}
0 & 1\\
0 & 0\end{array}\right],\\
B_3=\sqrt{\gamma_2}I_2\otimes\left[\begin{array}{cc}
1 & 0\\
0 & 0\end{array}\right],\\
B_4=\sqrt{1-\gamma_1-\gamma_2}I_2\otimes I_2.
\eeqan
 Any of the $B_i$ factorizes in an operator proportional to the identity on $\Hi_N$ times another on $\Hi_F$, this is a sufficient condition for $\Hi_N$ to be a \emph{Noiseless Subsystem} \cite{viola-generalnoise,ticozzi-isometries}.  Moreover in this decomposition projecting onto the the subspaces $\Hi_{S_1}$ and $\Hi_{S_2},$ defined above, correspond to projecting onto the states $\ket{0_N}$ and $\ket{1_N}$.  Thus evaluating the trace of the state projected onto one of them returns the probability of having prepared the corresponding state in $\Hi_N$. To do this in the asymptotic limit we can use the results of Section \ref{sec:asProb}, since both subspaces are invariant.

Turning to the asymptotic probabilities, it is convenient to evaluate the limits of the projections, as is done in the proof of Proposition \ref{prop:asymProb}, and then apply them to the initial state:
{\color{black} \footnotesize \begin{align*}
\lim\limits_{n\to\infty} T^{*n}(\Pi_{S_1})&=&\ket{1}\bra{1}+\ket{3}\bra{3}+\frac{1}{2}(\ket{5}\bra{5}+\ket{6}\bra{6}+\ket{7}\bra{7}),\\
\lim\limits_{n\to\infty} T^{*n}(\Pi_{S_2})&=&\ket{2}\bra{2}+\ket{4}\bra{4}+\frac{1}{2}(\ket{5}\bra{5}+\ket{6}\bra{6}+\ket{7}\bra{7}).
\end{align*} }
By these, if the initial state is $\rho_0 = \frac{1}{7}I_{7}$, we obtain:
\beqan
\lim\limits_{n\to\infty} \tr(\Pi_{S_1}T^n(\rho_0))=\frac{1}{2},\\
\lim\limits_{n\to\infty} \tr(\Pi_{S_2}T^n(\rho_0))=\frac{1}{2}.
\eeqan
If instead the initial state is $\rho_0 = \frac{1}{2}(\ket{1}\bra{1}+\ket{7}\bra{7})$, then we have
\beqan
\lim\limits_{n\to\infty} \tr(\Pi_{S_1}T^n(\rho_0))=\frac{3}{4},\\
\lim\limits_{n\to\infty} \tr(\Pi_{S_2}T^n(\rho_0))=\frac{1}{4}.
\eeqan

\section{Discussion and Conclusions}
A thorough understanding of open system dynamics, and specifically iterated CPTP maps, plays a central role in the development of quantum information and control methods. Their asymptotic behavior  reveals the effectiveness of error protection and correction strategies \cite{viola-IPSlong,ticozzi-isometries} as well as protocols for state preparation \cite{bolognani-arxiv}.

In this work we presented a rich set of linear-algebraic tools for analyzing the convergence features of a discrete-time QDS generated by a given CPTP map. The work complements known results on the decomposition of the peripheral eigen-operators of a CPTP map, and the emerging Hilbert-space structure \cite{wolf-notes,viola-IPS}, by introducing decompositions of the ``decaying'' part of the Hilbert space that highlight the mechanism and speed of convergence. Similarly to those results, our results build on a generalization of Perron-Frobenius theory. With respect to the existing results in continuous-time, the discrete-time is more general, and has direct applications in the study of engineered dynamics in quantum ``digital'' simulators \cite{barreiro}. 

{\ftr Our tools include two Hilbert space decompositions in ``transient'' subspaces. We believe that both have their place and potential advantages for certain tasks, which we next briefly discuss and compare. 

The nested-face decomposition is based on the positivity and spectral properties of the evolution, and its constructions is inspired by a Perron-Frobenius analysis. Its potential advantages include: (i) the ability of deciding GAS for the initial subspace at the first iteration -- in fact, if the first reduced spectral ratio is strictly lesser than one, convergence is guaranteed; (ii) the spectral radii correspond to exact asymptotic convergence speed; (iii) the nested subspaces $\Hi_{S_i}$ are all invariant subspaces, allowing for reduced description of the dynamics if needed. However, in the construction the initial representation of the dynamics may be soon lost, and is in general difficult to assess the role of a certain physical variable on the qualitative behavior.

On the other hand, the DID is based on the directed {\em dynamical links} between subspaces. Its construction decomposes the OSR matrices, and in practical examples where the maps are associated to physical decay processes (e.g. spontaneous or stimulated emissions), it often only requires a re-arranging of the natural basis. This can help in identifying the physical parameters leading to, or hindering, convergence. However, (i) to decide GAS one has to run the algorithm to its end; (ii) the transition rates are not exact asymptotic convergence speeds; (iii) its $\Hi_{S_i}$ are {\em not} invariant, with exception of the first one.

Summing up, the first one is more natural from a mathematical perspective, however the second is more likely to highlight the physical mechanisms leading to convergence \cite{ticozzi-NV}. }

Nonetheless, both decomposition associate subspaces with different decay speeds: eigenvalues of the map in the nested-face decomposition, or transition {\ic rates} in the DID. These speeds can help identifying bottlenecks for the convergence, typically associated to a particular physical process, as we illustrated in the example. 

{\ftr The asymptotic probability formula given in Proposition \ref{prop:asymProb} allows for computing the asymptotic probability distribution of converging to a set of orthogonal subspaces. This type of problems are relevant e.g. in verifying the efficiency of an initialization procedure for a quantum noiseless code, the probability of decoherence driving the state to one of a set of orthogonal pointer states, or in general to be able to assess the asymptotic properties when the evolution is not mixing, i.e. it does not have a unique GAS state.}

{\ft In conclusion, we derive a set of analysis tools that can aid in the design of evolution for quantum control and quantum information processing. In particular, they should provide suitable means} for analyzing the convergence speed of quantum information protocols based on dissipation, including entanglement preparation and computation \cite{wolf-dissipativeqc,kraus-entangled,ticozzi-QL,ticozzi-QIC}.

\acknowledgements
F.T. is pleased to thank Lorenza Viola for discussions on the material of this paper, and for the joint work that led to the formulation of the DID in the continuous-time case.
\appendix
\section{Proof of proposition~\ref{prop:decRad}}\label{ap:decRadProof}
The proof is based on two lemmas. The first one shows that an eigenvector for a reduced map on the complement of an invariant subspace can always be {\em extended} to a generalized eigenvector of the whole map corresponding to the same eigenvalue.

\begin{lemma}\label{lem:eigQuot}
Let $T$ be a linear map on a vector space ${\mathcal{V}=\mathcal{V}_1\oplus\mathcal{V}_2}$ of dimension $d$, with $\mathcal{V}_1$ invariant, so that 
\beq
T=\left[\begin{array}{cc}
T_1 & T_2 \\
0 & T_3
\end{array}\right].
\eeq
If $\eta \in \mathcal{V}_2$, $\eta=\left[\begin{array}{c}
0 \\
\eta_2
\end{array}\right]$
and $T_3 \eta_2 = \sigma\eta_2$, then exists $\xi \in \mathcal{V}_1$ such that
\beq
\label{eq:lhs}
\xi+\eta \in \ker((T-\sigma I)^{d}).
\eeq
\end{lemma}
\proof
If $\xi \in \mathcal{V}_1$ then
\[
\xi=\left[\begin{array}{c}
\xi_1 \\
0
\end{array}\right],
\]
and the lemma is proved if the equation in $\xi$
{\color{black}
\beq \label{eq:extendEq}
(T-\sigma I)^d(\xi+\eta)=\left[\begin{array}{c}
(T_1-\sigma I)^d \xi_1 + {T}_2 \eta_2 \\
(T_3-\sigma I)^d \eta_2
\end{array}\right] = 0
\eeq}%
has solution.

Since $\eta_2$ is an eigenvector of $T_3$, $(T_3-\sigma I)^d \eta_2=0$ and \eqref{eq:extendEq} reduces to
\beq 
(T_1-\sigma I)^d \xi_1 = -{T}_2 \eta_2.
\eeq
If $\sigma$ is not an eigenvalue of $T_1$, this system is clearly solvable in $\xi_1$.
In the other case it is solvable only if ${T}_2 \eta_2$ belongs to the image of $(T_1-\sigma I)^d$ i.e. if ${T}_2 \eta_2$ has no component with respect to the  generalized eigenspace relative to $\sigma$. Since generalized eigenspaces are in direct sum, ${T}_2 \eta_2$ admits a unique decomposition as sum of {\em generalized eigenvectors of $T_1$}:
\beq
{T}_2 \eta_2 = \sum\limits_k v_k,
\eeq
where any $v_k$ is relative to a different eigenvalue $\lambda_k$, with $\lambda_j \neq \lambda_i$ if $j\neq i$. Also, due to the invariance of $\mathcal{V}_1$, any $v_k$ can be (trivially) extended to a generalized eigenvector $w_k$ for $T,$ also relative to $\lambda_k$.
We can thus write:
\beq\label{eq:TeigDec}
(T-\sigma I)^d \eta=\left[\begin{array}{c}
{T}_2 \eta_2 \\
0
\end{array}\right] = \sum\limits_k \left[\begin{array}{c}
v_k \\
0
\end{array}\right] = \sum\limits_k w_k.
\eeq
Notice that if $\sigma$ is an eigenvalue of $T_1$, as it is in the case we are discussing, then it is also an eigenvalue of $T$ since $\mathcal{V}_1$ is invariant. It is thus apparent that none of the $w_k$ can be a generalized eigenvalue corresponding to $\sigma$: in fact, due to the Jordan structure of $T$, $T-\sigma I$ restricted to the generalized $\sigma$-eigenspace is a nilpotent matrix of order at most $d$, and hence any generalized $\sigma$-eigenvector for $T$ is mapped to zero after $d$ applications of $T-\sigma I$.
This allows us to conclude that none of the $v_k$ is relative to $\sigma$ and the system \eqref{eq:extendEq} is always solvable.
\qed
\begin{lemma}\label{lem:pos}
Let $\Hi$ be a finite dimensional Hilbert space, and $\Hi_S\oplus\Hi_R$ an orthogonal decomposition.\\
If $Z \in \HH$ has block form
\beq
Z=\left[\begin{array}{cc}
Z_S & Z_{P} \\
Z_{P}^\dagger & Z_R
\end{array}\right],
\eeq
with $Z_R>0$ and $X \in \HH$ is such that
\beq
X=\left[\begin{array}{cc}
X_S & 0 \\
0 & 0
\end{array}\right],
\eeq
with $X_S >0$, then there exists a scalar $c>0$ such that $Z+cX > 0$.
\end{lemma}
\proof
Considering the block form induced by the decomposition of $\Hi$, if $\ket{\varphi} = \ket{\varphi_S}\oplus\ket{\varphi_R}$ then

\makeatletter
\renewcommand{\maketag@@@}[1]{\hbox{\m@th\normalsize\normalfont#1}}%
\makeatother

{\color{black} \footnotesize
\begin{gather}
\bra{\varphi}Z+cX\ket{\varphi} =
\bra{\varphi_S}\oplus\bra{\varphi_R}\left[\begin{array}{cc}
Z_S+cX_S & Z_{P}  \\
 Z_{P}^\dagger & Z_R
\end{array}\right]\ket{\varphi_S}\oplus\ket{\varphi_R}=\nonumber \\
\bra{\varphi_S} cX_S + Z_S\ket{\varphi_S}+ \bra{\varphi_R} Z_{P}^\dagger\ket{\varphi_S} + \bra{\varphi_S} Z_{P} \ket{\varphi_R} + \bra{\varphi_R }Z_{R}\ket{\varphi_R}. \label{eq:lemPos5} 
\end{gather} }
Since $X_S>0$ exists a $c_1$ such that $c_1X_S+Z_S > 0$, so that redefining $c=c'+c_1$:
\beq\label{eq:lemPos2}
\bra{ \varphi_S} cX_S + Z_S\ket{\varphi_S}  \geq \bra{ \varphi_S} c'X_S \ket{\varphi_S}
\eeq
for any $\ket{\varphi_S}$.\\
The set of vectors $\ket{\varphi_S} \oplus \ket{\varphi_R}$, under the condition $\braket{\varphi_S}{\varphi_S}=\braket{\varphi_R}{\varphi_R}=1$, is compact and $\bra{\varphi_R}Z_{P}^\dagger\ket{\varphi_S} + \bra{\varphi_S}Z_{P}\ket{\varphi_R}$ is a real continuous function, then  exists $m > 0 $ such that
\beq
\langle\varphi_R |Z_{P}^\dagger|\varphi_S\rangle + \langle\varphi_S|Z_{P}|\varphi_R\rangle \geq -2m,
\eeq
if $|\varphi_S|=|\varphi_R|=1$. So for $\ket{\varphi_S}$ and $\ket{\varphi_R }$ with $|\varphi_S|\neq 0$ and $|\varphi_R|\neq 0$ 
{\color{black} \footnotesize
\begin{gather}
\bra{\varphi_R} Z_{P}^\dagger \ket{\varphi_S}+ \bra{\varphi_S }Z_{P}\ket{\varphi_R}=\nonumber \\
|\varphi_R||\varphi_S|\left(\bra{ \frac{1}{|\varphi_R|}\varphi_R }Z_{P}^\dagger\ket{\frac{1}{|\varphi_S|}\varphi_S} + \bra{\frac{1}{|\varphi_S|}\varphi_S }Z_{P}\ket{\frac{1}{|\varphi_R|}\varphi_R}\right)\geq \nonumber \\ -2m|\varphi_R||\varphi_S|.\label{eq:lemPos3}
\end{gather} }
By positiveness of $X_S$ and $Z_R$
\beq\label{eq:lemPos4}
\begin{aligned}
\bra{\varphi_S} X_S\ket{\varphi_S}\ \geq a_1^2 |\varphi_S|^2 \\
\bra{\varphi_R} Z_R \ket{\varphi_R}\ \geq a_2^2 |\varphi_R|^2 
\end{aligned}
\eeq
for some $a_1 \, , a_2 > 0$ (for example the roots of their minimum eigenvalues).\\
Suppose $|\varphi_S|\neq 0$ and $|\varphi_R|\neq 0$ and put \eqref{eq:lemPos2}, \eqref{eq:lemPos3} and \eqref{eq:lemPos4} in \eqref{eq:lemPos5}
\beq
\begin{aligned}
&\bra{ \varphi_S } \oplus \bra{ \varphi_R} cX+Z \ket{\varphi_S} \oplus \ket{\varphi_R}\geq\\
&a_1^2 |\varphi_S|^2 c'+a_2^2|\varphi_R|^2-2m|\varphi_S||\varphi_R|
\end{aligned}
\eeq
choose $c'> \left(\frac{m}{a_1 a_2}\right)^2$
\beq
\begin{aligned}
&\bra{\varphi_S} \oplus \bra{ \varphi_R} cX+Z \ket{\varphi_S} \oplus \ket{\varphi_R}>\\
&\frac{m^2}{a_2^2}|\varphi_S|^2 +a_2^2|\varphi_R|^2-2m|\varphi_S||\varphi_R|=\\
&(\frac{m}{a_2}|\varphi_S|-a_2 |\varphi_R|)^2 > 0,
\end{aligned}
\eeq
so $cX+Z > 0$.
\qed
{\ic It is now possible to prove proposition~\ref{prop:decRad}.}

\vspace{2mm}{\em Proof of proposition \ref{prop:decRad}.}
Suppose by contradiction that $\sigma=\sigma(T_{R_i})=\sigma(T_{R_{i+1}})$. It would be then possible to find $A\geq 0$ such that $T_{R_{i+1}}(A)=\sigma A$. $\Hi_T'=\supp(A)$ is invariant for  $T_{R_{i+1}}$ so that $\Hi_R'=\Hi_{T_i}\oplus \Hi_T'$ is invariant for $T_{R_i}$. Consider $T_R'$ the restriction of $T_{R_i}$ to $\Hi_R'$. For this map $\HHs{T_i}\oplus \HHs{T_i T'}$ is invariant and, being $A \in \HHs{T'}$ is an eigenoperator for $T_{T'}$,  it is thus possible to apply Lemma \ref{lem:eigQuot} to extend $A$ to a generalized eigenoperator of $T_R'$, of the form
\beq
A'=\left[\begin{array}{cc}
A_1 & A_2\\
A_2^\dagger & A
\end{array}\right],
\eeq
with $A>0$. By the definition of $\Hi_{T_i}$ there exists, for $T_{R_i}$, a generalized eigenvector $X\geq 0$ relative to $\sigma$, such that $\supp(X)=\Hi_{T_i}$. Since $\Hi_{T_i} \subset \Hi_R'$ $X$ is an eigenvalue of $T_R'$ as well. Now by Lemma \ref{lem:pos} it is possible to find a constant $c>0$ such that $B=cX+A>0$. Since $B$ is a generalized eigenvector of $T_R'$  relative to $\sigma$, the same holds for $T_{R_i}$. However, this is not possible since $\Hi_{T_i}$ is strictly contained in $\supp(B),$ in contradiction with its definition.
\qed

\bibliography{bib-discrete-DID}

\begin{thebibliography}{33}
\expandafter\ifx\csname natexlab\endcsname\relax\def\natexlab#1{#1}\fi
\expandafter\ifx\csname bibnamefont\endcsname\relax
  \def\bibnamefont#1{#1}\fi
\expandafter\ifx\csname bibfnamefont\endcsname\relax
  \def\bibfnamefont#1{#1}\fi
\expandafter\ifx\csname citenamefont\endcsname\relax
  \def\citenamefont#1{#1}\fi
\expandafter\ifx\csname url\endcsname\relax
  \def\url#1{\texttt{#1}}\fi
\expandafter\ifx\csname urlprefix\endcsname\relax\def\urlprefix{URL }\fi
\providecommand{\bibinfo}[2]{#2}
\providecommand{\eprint}[2][]{\url{#2}}

\bibitem[{\citenamefont{Davies}(1976)}]{davies}
\bibinfo{author}{\bibfnamefont{E.~B.} \bibnamefont{Davies}},
  \emph{\bibinfo{title}{Quantum Theory of Open Systems}}
  (\bibinfo{publisher}{Academic Press, USA}, \bibinfo{year}{1976}).

\bibitem[{\citenamefont{Kraus}(1983)}]{kraus}
\bibinfo{author}{\bibfnamefont{K.}~\bibnamefont{Kraus}},
  \emph{\bibinfo{title}{States, Effects, and Operations: Fundamental Notions of
  Quantaum Theory}}, Lecture notes in Physics
  (\bibinfo{publisher}{Springer-Verlag, Berlin}, \bibinfo{year}{1983}).

\bibitem[{\citenamefont{Bratteli and Robinson}(1979)}]{bratteli}
\bibinfo{author}{\bibfnamefont{O.}~\bibnamefont{Bratteli}} \bibnamefont{and}
  \bibinfo{author}{\bibfnamefont{D.~W.} \bibnamefont{Robinson}},
  \emph{\bibinfo{title}{Operator Algebras and Quantum Statistical Mechanics,
  vol. I {\&} II}} (\bibinfo{publisher}{Springer-Verlag, Berlin},
  \bibinfo{year}{1979}).

\bibitem[{\citenamefont{Nielsen and Chuang}(2002)}]{nielsen-chuang}
\bibinfo{author}{\bibfnamefont{M.~A.} \bibnamefont{Nielsen}} \bibnamefont{and}
  \bibinfo{author}{\bibfnamefont{I.~L.} \bibnamefont{Chuang}},
  \emph{\bibinfo{title}{Quantum Computation and Information}}
  (\bibinfo{publisher}{Cambridge University Press, Cambridge},
  \bibinfo{year}{2002}).

\bibitem[{\citenamefont{Altafini and Ticozzi}(2012)}]{altafini-tutorial}
\bibinfo{author}{\bibfnamefont{C.}~\bibnamefont{Altafini}} \bibnamefont{and}
  \bibinfo{author}{\bibfnamefont{F.}~\bibnamefont{Ticozzi}},
  \bibinfo{journal}{IEEE Trans. Aut. Control} \textbf{\bibinfo{volume}{57}},
  \bibinfo{pages}{1898} (\bibinfo{year}{2012}).

\bibitem[{\citenamefont{Zurek}(2003)}]{zurek-decoherence}
\bibinfo{author}{\bibfnamefont{W.~H.} \bibnamefont{Zurek}},
  \bibinfo{journal}{Rev. Mod. Phys.} \textbf{\bibinfo{volume}{75}},
  \bibinfo{pages}{715} (\bibinfo{year}{2003}),
  \urlprefix\url{http://link.aps.org/doi/10.1103/RevModPhys.75.715}.

\bibitem[{\citenamefont{Burgarth et~al.}(2013)\citenamefont{Burgarth,
  Chiribella, Giovannetti, Perinotti, and Yuasa}}]{burgarth-mixing}
\bibinfo{author}{\bibfnamefont{D.}~\bibnamefont{Burgarth}},
  \bibinfo{author}{\bibfnamefont{G.}~\bibnamefont{Chiribella}},
  \bibinfo{author}{\bibfnamefont{V.}~\bibnamefont{Giovannetti}},
  \bibinfo{author}{\bibfnamefont{P.}~\bibnamefont{Perinotti}},
  \bibnamefont{and} \bibinfo{author}{\bibfnamefont{K.}~\bibnamefont{Yuasa}},
  \bibinfo{journal}{New Journal of Physics} \textbf{\bibinfo{volume}{15}},
  \bibinfo{pages}{073045} (\bibinfo{year}{2013}).

\bibitem[{\citenamefont{Kastoryano et~al.}(2012)\citenamefont{Kastoryano, Reeb,
  and Wolf}}]{reeb-cutoff}
\bibinfo{author}{\bibfnamefont{M.~J.} \bibnamefont{Kastoryano}},
  \bibinfo{author}{\bibfnamefont{D.}~\bibnamefont{Reeb}}, \bibnamefont{and}
  \bibinfo{author}{\bibfnamefont{M.~M.} \bibnamefont{Wolf}},
  \bibinfo{journal}{Journal of Physics A: Mathematical and Theoretical}
  \textbf{\bibinfo{volume}{45}}, \bibinfo{pages}{075307}
  (\bibinfo{year}{2012}).

\bibitem[{\citenamefont{Lidar et~al.}(1997)\citenamefont{Lidar, Chuang, and
  Whaley}}]{lidar-DFS}
\bibinfo{author}{\bibfnamefont{D.~A.} \bibnamefont{Lidar}},
  \bibinfo{author}{\bibfnamefont{I.~L.} \bibnamefont{Chuang}},
  \bibnamefont{and} \bibinfo{author}{\bibfnamefont{K.~B.}
  \bibnamefont{Whaley}}, \bibinfo{journal}{Phys. Rev. Lett.}
  \textbf{\bibinfo{volume}{81}}, \bibinfo{pages}{2594} (\bibinfo{year}{1997}).

\bibitem[{\citenamefont{Ticozzi and Viola}(2008)}]{ticozzi-QDS}
\bibinfo{author}{\bibfnamefont{F.}~\bibnamefont{Ticozzi}} \bibnamefont{and}
  \bibinfo{author}{\bibfnamefont{L.}~\bibnamefont{Viola}},
  \bibinfo{journal}{IEEE Trans. Aut. Contr.} \textbf{\bibinfo{volume}{53}},
  \bibinfo{pages}{2048} (\bibinfo{year}{2008}).

\bibitem[{\citenamefont{Ticozzi and Viola}(2009)}]{ticozzi-markovian}
\bibinfo{author}{\bibfnamefont{F.}~\bibnamefont{Ticozzi}} \bibnamefont{and}
  \bibinfo{author}{\bibfnamefont{L.}~\bibnamefont{Viola}},
  \bibinfo{journal}{Automatica} \textbf{\bibinfo{volume}{45}},
  \bibinfo{pages}{2002} (\bibinfo{year}{2009}).

\bibitem[{\citenamefont{Bolognani and Ticozzi}(2010)}]{bolognani-arxiv}
\bibinfo{author}{\bibfnamefont{S.}~\bibnamefont{Bolognani}} \bibnamefont{and}
  \bibinfo{author}{\bibfnamefont{F.}~\bibnamefont{Ticozzi}},
  \bibinfo{journal}{IEEE Trans. Aut. Contr.} \textbf{\bibinfo{volume}{55}},
  \bibinfo{pages}{2721 } (\bibinfo{year}{2010}), ISSN
  \bibinfo{issn}{0018-9286}.

\bibitem[{\citenamefont{Ticozzi et~al.}(2013)\citenamefont{Ticozzi, Nishio, and
  Altafini}}]{ticozzi-stochastic}
\bibinfo{author}{\bibfnamefont{F.}~\bibnamefont{Ticozzi}},
  \bibinfo{author}{\bibfnamefont{K.}~\bibnamefont{Nishio}}, \bibnamefont{and}
  \bibinfo{author}{\bibfnamefont{C.}~\bibnamefont{Altafini}},
  \bibinfo{journal}{Automatic Control, IEEE Transactions on}
  \textbf{\bibinfo{volume}{58}}, \bibinfo{pages}{74} (\bibinfo{year}{2013}),
  ISSN \bibinfo{issn}{0018-9286}.

\bibitem[{\citenamefont{Mirrahimi and Handel}(2007)}]{mirrahimi-stabilization}
\bibinfo{author}{\bibfnamefont{M.}~\bibnamefont{Mirrahimi}} \bibnamefont{and}
  \bibinfo{author}{\bibfnamefont{R.~V.} \bibnamefont{Handel}},
  \bibinfo{journal}{SIAM J. Control Optim.} \textbf{\bibinfo{volume}{46}},
  \bibinfo{pages}{445} (\bibinfo{year}{2007}), ISSN \bibinfo{issn}{0363-0129}.

\bibitem[{\citenamefont{Albertini and Ticozzi}(2011)}]{albertini-feedback}
\bibinfo{author}{\bibfnamefont{F.}~\bibnamefont{Albertini}} \bibnamefont{and}
  \bibinfo{author}{\bibfnamefont{F.}~\bibnamefont{Ticozzi}},
  \bibinfo{journal}{Automatica} \textbf{\bibinfo{volume}{47}},
  \bibinfo{pages}{2451 } (\bibinfo{year}{2011}).

\bibitem[{\citenamefont{Yamamoto et~al.}(2007)\citenamefont{Yamamoto, Tsumura,
  and Hara}}]{yamamoto-twospin}
\bibinfo{author}{\bibfnamefont{N.}~\bibnamefont{Yamamoto}},
  \bibinfo{author}{\bibfnamefont{K.}~\bibnamefont{Tsumura}}, \bibnamefont{and}
  \bibinfo{author}{\bibfnamefont{S.}~\bibnamefont{Hara}},
  \bibinfo{journal}{Automatica} \textbf{\bibinfo{volume}{43}},
  \bibinfo{pages}{981} (\bibinfo{year}{2007}).

\bibitem[{\citenamefont{Ticozzi and Viola}(2012)}]{ticozzi-QL}
\bibinfo{author}{\bibfnamefont{F.}~\bibnamefont{Ticozzi}} \bibnamefont{and}
  \bibinfo{author}{\bibfnamefont{L.}~\bibnamefont{Viola}},
  \bibinfo{journal}{Phil. Trans. R. Soc. A} \textbf{\bibinfo{volume}{370}},
  \bibinfo{pages}{5259} (\bibinfo{year}{2012}).

\bibitem[{\citenamefont{Ticozzi and Viola}(2014)}]{ticozzi-QIC}
\bibinfo{author}{\bibfnamefont{F.}~\bibnamefont{Ticozzi}} \bibnamefont{and}
  \bibinfo{author}{\bibfnamefont{L.}~\bibnamefont{Viola}},
  \bibinfo{journal}{Quantum Inf. Comput.} \textbf{\bibinfo{volume}{14}},
  \bibinfo{pages}{0265} (\bibinfo{year}{2014}).

\bibitem[{\citenamefont{Barreiro et~al.}(2011)\citenamefont{Barreiro,
  M\"{u}ller, Schindler, Nigg, Monz, Chwalla, Hennrich, Roos, Zoller, and
  Blatt}}]{barreiro}
\bibinfo{author}{\bibfnamefont{J.~T.} \bibnamefont{Barreiro}},
  \bibinfo{author}{\bibfnamefont{M.}~\bibnamefont{M\"{u}ller}},
  \bibinfo{author}{\bibfnamefont{P.}~\bibnamefont{Schindler}},
  \bibinfo{author}{\bibfnamefont{D.}~\bibnamefont{Nigg}},
  \bibinfo{author}{\bibfnamefont{T.}~\bibnamefont{Monz}},
  \bibinfo{author}{\bibfnamefont{M.}~\bibnamefont{Chwalla}},
  \bibinfo{author}{\bibfnamefont{M.}~\bibnamefont{Hennrich}},
  \bibinfo{author}{\bibfnamefont{C.~F.} \bibnamefont{Roos}},
  \bibinfo{author}{\bibfnamefont{P.}~\bibnamefont{Zoller}}, \bibnamefont{and}
  \bibinfo{author}{\bibfnamefont{R.}~\bibnamefont{Blatt}},
  \bibinfo{journal}{Nature} \textbf{\bibinfo{volume}{470}},
  \bibinfo{pages}{486} (\bibinfo{year}{2011}).

\bibitem[{\citenamefont{Schindler et~al.}(2013)\citenamefont{Schindler,
  M\"{u}ller, Nigg, Barreiro, Martinez, Hennrich, Monz, Diehl, Zoller, and
  Blatt}}]{barreiro2}
\bibinfo{author}{\bibfnamefont{P.}~\bibnamefont{Schindler}},
  \bibinfo{author}{\bibfnamefont{M.}~\bibnamefont{M\"{u}ller}},
  \bibinfo{author}{\bibfnamefont{D.}~\bibnamefont{Nigg}},
  \bibinfo{author}{\bibfnamefont{J.~T.} \bibnamefont{Barreiro}},
  \bibinfo{author}{\bibfnamefont{E.~A.} \bibnamefont{Martinez}},
  \bibinfo{author}{\bibfnamefont{M.}~\bibnamefont{Hennrich}},
  \bibinfo{author}{\bibfnamefont{T.}~\bibnamefont{Monz}},
  \bibinfo{author}{\bibfnamefont{S.}~\bibnamefont{Diehl}},
  \bibinfo{author}{\bibfnamefont{P.}~\bibnamefont{Zoller}}, \bibnamefont{and}
  \bibinfo{author}{\bibfnamefont{R.}~\bibnamefont{Blatt}},
  \bibinfo{journal}{Nature Phys.} \textbf{\bibinfo{volume}{9}},
  \bibinfo{pages}{361} (\bibinfo{year}{2013}).

\bibitem[{\citenamefont{Alicki and Lendi}(1987)}]{alicki-lendi}
\bibinfo{author}{\bibfnamefont{R.}~\bibnamefont{Alicki}} \bibnamefont{and}
  \bibinfo{author}{\bibfnamefont{K.}~\bibnamefont{Lendi}},
  \emph{\bibinfo{title}{Quantum Dynamical Semigroups and Applications}}
  (\bibinfo{publisher}{Springer-Verlag, Berlin}, \bibinfo{year}{1987}).

\bibitem[{\citenamefont{Baumgartner et~al.}(2008)\citenamefont{Baumgartner,
  Narnhofer, and Thirring}}]{baumgartner-1}
\bibinfo{author}{\bibfnamefont{B.}~\bibnamefont{Baumgartner}},
  \bibinfo{author}{\bibfnamefont{H.}~\bibnamefont{Narnhofer}},
  \bibnamefont{and} \bibinfo{author}{\bibfnamefont{W.}~\bibnamefont{Thirring}},
  \bibinfo{journal}{Journal of Physics A: Mathematical and Theoretical}
  \textbf{\bibinfo{volume}{41}}, \bibinfo{pages}{065201}
  (\bibinfo{year}{2008}).

\bibitem[{\citenamefont{Baumgartner and Narnhofer}(2008)}]{baumgartner-2}
\bibinfo{author}{\bibfnamefont{B.}~\bibnamefont{Baumgartner}} \bibnamefont{and}
  \bibinfo{author}{\bibfnamefont{H.}~\bibnamefont{Narnhofer}},
  \bibinfo{journal}{Journal of Physics A: Mathematical and Theoretical}
  \textbf{\bibinfo{volume}{41}}, \bibinfo{pages}{395303}
  (\bibinfo{year}{2008}).

\bibitem[{\citenamefont{Ticozzi et~al.}(2012)\citenamefont{Ticozzi, Lucchese,
  Cappellaro, and Viola}}]{ticozzi-NV}
\bibinfo{author}{\bibfnamefont{F.}~\bibnamefont{Ticozzi}},
  \bibinfo{author}{\bibfnamefont{R.}~\bibnamefont{Lucchese}},
  \bibinfo{author}{\bibfnamefont{P.}~\bibnamefont{Cappellaro}},
  \bibnamefont{and} \bibinfo{author}{\bibfnamefont{L.}~\bibnamefont{Viola}},
  \bibinfo{journal}{IEEE Transactions on Automatic Control}
  \textbf{\bibinfo{volume}{57}}, \bibinfo{pages}{1931} (\bibinfo{year}{2012}).

\bibitem[{\citenamefont{Wolf and Cirac}(2008)}]{wolf-dividing}
\bibinfo{author}{\bibfnamefont{M.}~\bibnamefont{Wolf}} \bibnamefont{and}
  \bibinfo{author}{\bibfnamefont{J.}~\bibnamefont{Cirac}},
  \bibinfo{journal}{Communications in Mathematical Physics}
  \textbf{\bibinfo{volume}{279}}, \bibinfo{pages}{147} (\bibinfo{year}{2008}).

\bibitem[{\citenamefont{Wolf}(2012)}]{wolf-notes}
\bibinfo{author}{\bibfnamefont{M.~M.} \bibnamefont{Wolf}},
  \emph{\bibinfo{title}{Quantum Channels \& Operations: A Guided Tour}}
  (\bibinfo{publisher}{Lecture notes available at\\
  http://www-m5.ma.tum.de/foswiki/pub/M5\\/Allgemeines/MichaelWolf/QChannelLecture.pdf},
  \bibinfo{year}{2012}).

\bibitem[{\citenamefont{Hill and Waters}(1987)}]{hill1987cone}
\bibinfo{author}{\bibfnamefont{R.~D.} \bibnamefont{Hill}} \bibnamefont{and}
  \bibinfo{author}{\bibfnamefont{S.~R.} \bibnamefont{Waters}},
  \bibinfo{journal}{Linear Algebra and its Applications}
  \textbf{\bibinfo{volume}{90}}, \bibinfo{pages}{81} (\bibinfo{year}{1987}).

\bibitem[{\citenamefont{Ticozzi and Viola}(2010)}]{ticozzi-isometries}
\bibinfo{author}{\bibfnamefont{F.}~\bibnamefont{Ticozzi}} \bibnamefont{and}
  \bibinfo{author}{\bibfnamefont{L.}~\bibnamefont{Viola}},
  \bibinfo{journal}{Phys. Rev. A} \textbf{\bibinfo{volume}{81}},
  \bibinfo{pages}{032313} (\bibinfo{year}{2010}).

\bibitem[{\citenamefont{Knill et~al.}(2000)\citenamefont{Knill, Laflamme, and
  Viola}}]{viola-generalnoise}
\bibinfo{author}{\bibfnamefont{E.}~\bibnamefont{Knill}},
  \bibinfo{author}{\bibfnamefont{R.}~\bibnamefont{Laflamme}}, \bibnamefont{and}
  \bibinfo{author}{\bibfnamefont{L.}~\bibnamefont{Viola}},
  \bibinfo{journal}{Phys. Rev. Lett.} \textbf{\bibinfo{volume}{84}},
  \bibinfo{pages}{2525} (\bibinfo{year}{2000}).

\bibitem[{\citenamefont{Blume-Kohout et~al.}(2010)\citenamefont{Blume-Kohout,
  Ng, Poulin, and Viola}}]{viola-IPSlong}
\bibinfo{author}{\bibfnamefont{R.}~\bibnamefont{Blume-Kohout}},
  \bibinfo{author}{\bibfnamefont{H.~K.} \bibnamefont{Ng}},
  \bibinfo{author}{\bibfnamefont{D.}~\bibnamefont{Poulin}}, \bibnamefont{and}
  \bibinfo{author}{\bibfnamefont{L.}~\bibnamefont{Viola}},
  \bibinfo{journal}{Phys. Rev. A} \textbf{\bibinfo{volume}{82}},
  \bibinfo{pages}{062306} (\bibinfo{year}{2010}).

\bibitem[{\citenamefont{Blume-Kohout et~al.}(2008)\citenamefont{Blume-Kohout,
  Ng, Poulin, and Viola}}]{viola-IPS}
\bibinfo{author}{\bibfnamefont{R.}~\bibnamefont{Blume-Kohout}},
  \bibinfo{author}{\bibfnamefont{H.~K.} \bibnamefont{Ng}},
  \bibinfo{author}{\bibfnamefont{D.}~\bibnamefont{Poulin}}, \bibnamefont{and}
  \bibinfo{author}{\bibfnamefont{L.}~\bibnamefont{Viola}},
  \bibinfo{journal}{Phys. Rev. Lett.} \textbf{\bibinfo{volume}{100}},
  \bibinfo{pages}{030501} (\bibinfo{year}{2008}).

\bibitem[{\citenamefont{Verstraete et~al.}(2009)\citenamefont{Verstraete, Wolf,
  and Cirac}}]{wolf-dissipativeqc}
\bibinfo{author}{\bibfnamefont{F.}~\bibnamefont{Verstraete}},
  \bibinfo{author}{\bibfnamefont{M.~M.} \bibnamefont{Wolf}}, \bibnamefont{and}
  \bibinfo{author}{\bibfnamefont{J.~I.} \bibnamefont{Cirac}},
  \bibinfo{journal}{Nature Physics} \textbf{\bibinfo{volume}{5}},
  \bibinfo{pages}{633 } (\bibinfo{year}{2009}).

\bibitem[{\citenamefont{Kraus et~al.}(2008)\citenamefont{Kraus, Diehl, Micheli,
  Kantian, B\"{u}chler, and Zoller}}]{kraus-entangled}
\bibinfo{author}{\bibfnamefont{B.}~\bibnamefont{Kraus}},
  \bibinfo{author}{\bibfnamefont{S.}~\bibnamefont{Diehl}},
  \bibinfo{author}{\bibfnamefont{A.}~\bibnamefont{Micheli}},
  \bibinfo{author}{\bibfnamefont{A.}~\bibnamefont{Kantian}},
  \bibinfo{author}{\bibfnamefont{H.~P.} \bibnamefont{B\"{u}chler}},
  \bibnamefont{and} \bibinfo{author}{\bibfnamefont{P.}~\bibnamefont{Zoller}},
  \bibinfo{journal}{Phys. Rev. A} \textbf{\bibinfo{volume}{78}},
  \bibinfo{pages}{042307} (\bibinfo{year}{2008}).

\end{thebibliography}

\end{document}